\documentclass[11pt]{article}
\usepackage{changepage}
\usepackage[normalem]{ulem}

\usepackage{lineno} 

\usepackage{fullpage}
\usepackage[authoryear]{natbib}
\usepackage{xr-hyper}
\usepackage{hyperref}
\usepackage{amssymb,amsfonts,amsmath}
\usepackage{helvet}
\usepackage{amsmath}
\usepackage{amssymb}
\usepackage{graphicx}
\usepackage{color, xcolor} 
\usepackage{enumerate}
\usepackage{color,soul}
\newcommand{\blank}[1]{}

\blank{\usepackage{lineno}
\linenumbers
\newcommand*\patchAmsMathEnvironmentForLineno[1]{%
  \expandafter\let\csname old#1\expandafter\endcsname\csname #1\endcsname
  \expandafter\let\csname oldend#1\expandafter\endcsname\csname end#1\endcsname
  \renewenvironment{#1}%
     {\linenomath\csname old#1\endcsname}%
     {\csname oldend#1\endcsname\endlinenomath}}%
\newcommand*\patchBothAmsMathEnvironmentsForLineno[1]{%
  \patchAmsMathEnvironmentForLineno{#1}%
  \patchAmsMathEnvironmentForLineno{#1*}}%
\AtBeginDocument{%
\patchBothAmsMathEnvironmentsForLineno{equation}%
\patchBothAmsMathEnvironmentsForLineno{align}%
\patchBothAmsMathEnvironmentsForLineno{flalign}%
\patchBothAmsMathEnvironmentsForLineno{alignat}%
\patchBothAmsMathEnvironmentsForLineno{gather}%
\patchBothAmsMathEnvironmentsForLineno{multline}%
}
}

\usepackage{verbatim}
\usepackage[dvipsnames]{xcolor}
\usepackage{array}
\usepackage{arydshln}
\usepackage{adjustbox}
\usepackage{xfrac}

\newcommand{\Aa}{\mathbf{A}}        
\newcommand{\Rr}{\mathbf{R}}  
  
\newcommand{\Bb}{\mathbf{B}}  
\newcommand{\Bind}{\mathbf{B}_{ind}}  
\newcommand{\Jj}{\mathbf{J}} 
 
\newcommand{\Js}{\Jj^*}

\newcommand{\Id}{\mathbf{I}}

\newcommand{\diag}[1]{\text{diag}({#1})}

\newcommand{\cov}[1]{\text{Cov}(#1)}
\newcommand{\bardown}[2]{\left. #1 \right|_{#2}}

\title{Heterogeneity among and competition between life stages together extend the stability boundary of complex ecological communities}

\begin{document}

\maketitle

Grace Garmire $^{1}$
James P. O'Dwyer  $^{2}$
\\
\textbf{1} Department of Physics, University of Illinois Urbana-Champaign, Urbana IL USA \\
\textbf{2} Department of Plant Biology, University of Illinois Urbana-Champaign, Urbana IL USA \\

\begin{abstract}
\noindent 
Classical models of pairwise species interactions consistently predict that large, complex communities are inherently unstable, yet the observed persistence of diverse communities highlights a central contradiction in theoretical ecology. While foundational frameworks often model species as homogeneous populations, one theoretical resolution lies in including internal species structure. In this work, we incorporate explicit demographic stage structure into a generalized Lotka-Volterra framework to evaluate how ontogenetic changes throughout the life cycle of a species can affect whole-community stability. We show that stage-dependence of inter-specific interactions can significantly expand the stable parameter space for a species network. Furthermore, this enhanced stability requires strong competition between life stages of a species. Our study illustrates that stage structure with population heterogeneity expands classical stability criteria, providing a mathematical foundation for understanding how ontogeny modifies the classically derived complexity boundaries for local asymptotic stability.
\end{abstract}

\pagebreak

\section*{Introduction}

The relationship between the complexity of an ecological community and its stability has remained a central focus in theoretical ecology for decades. Resolving this relationship is key to reconciling mathematical theory with the persistence of highly diverse natural communities, and can provide the foundation for predicting how ecosystems will respond to accelerating biodiversity loss \citep{cowie2022}. Historically, ecological intuition suggested that greater diversity increased community stability, as systems with more species possess a larger number of pathways for energy flow, buffering the community against perturbations \citep{Macarthur_1955, elton1958}. To formalize these dynamics, ecologists often model populations of interacting species using systems of ordinary differential equations (ODEs), where stability is assessed by the system's return to or divergence from a feasible equilibrium following a perturbation \citep{allesina2015, davis2022}. However, accurately modeling specific dynamics is an extremely complex task, especially in  communities with a large number of species and many varied behaviors \citep{thebault2010, terry2017}. This requires exact knowledge of individual species behaviors and interactions, which are essentially impossible to empirically determine due to observational limitations \citep{pocock2021}. 

To bypass these empirical limitations, \citet{may1972} took an alternative approach, using random matrix theory to demonstrate a mathematical limit for community stability. Extending the work of \citet{gardner1970} and modeling community dynamics with randomly distributed interaction strengths, May showed that large, complex communities are inherently unstable \citep{may1972, may1973}. Yet the biological reality of this transition is nuanced: while experimental work on microbial communities provides empirical evidence of May's instability transition \textit{in vitro}, \citep{fox2002, ratzke2020, hu2022} highly diverse communities continue to persist despite changing conditions \citep{McCann2000, donohue2016}. This suggests that rather than stability being independent of complexity,  real communities possess structural mechanisms that can sustain a higher number of species than purely unstructured models predict. Explored structures include network organization with different interaction types \citep{Allesina_2012}, explicit consumer resource dynamics \citep{butler2018, butler2020, Gibbs2022}, mechanistic trait-based approaches \citep{loreau2013}, evolving network structures \citep{loeuille2005} and dynamic foraging adaptations \citep{Kondoh2003}.

A critical gap in the study of these stabilizing mechanisms is the frequent assumption in large-scale community modeling that each species is a homogeneous group, exhibiting identical behaviors and interactions across all individuals \citep{werner1984, vindenes2015}.
In biological reality, most species grow and change across their life stages, resulting in distinct behaviors, reproductive rates, sizes, and masses over an individual's lifespan \citep{werner1984, nakazawa2011, Ke2018}. Within a single species, these developmental stages interact explicitly through the life cycle: juveniles mature into older stages, and adults reproduce to generate new offspring. Interactions with other species can also change between stages, where juveniles and adults may experience entirely different predation rates and competitive pressures \citep{hamrin1986, bystrom2005}. Empirical studies have also illustrated how ontogenetic shifts during a species' lifespan modify food web structure \citep{clegg2018}. 

Age- or stage-structured models are widely utilized in general population ecology, but their application to studies of the stability of large, complex communities has been limited, in part because of the increased complexity it brings \citep{wangersky1978}. Previous work has established that stage structure can fundamentally alter food web dynamics \citep{nilsson2018}. Notably, \citet{deRoos2021} demonstrated that accounting for intra-specific juvenile-adult dynamics can stabilize large systems if the rates governing reproduction and growth respond dynamically to changing resource availability. Building on this, \citet{gimenez2025} showed that when cross-stage interactions are biased toward specific dynamic types, such as predator-prey or competition, they can have either stabilizing or destabilizing effects. Furthermore, broader studies highlight how stage structure inherently alters network complexity \citep{rudolf2011} and introduces symmetry breaking into ecological models \citep{persson2013}. 

While these studies establish the importance of including stage structure in population models, they largely focus on specific network architectures or resource-dependent responses. The explicit mathematical consequences of stage-dependent heterogeneity in the interactions, or functional diversity, between stages on the local stability of large, complex generalized Lotka-Volterra systems remains underexplored. To address this gap, we modify the generalized Lotka-Volterra framework to include multiple life stages for each species. In this model, population dynamics emerge both from life stages interacting via a species life cycle and explicitly stage-dependent inter-specific interactions. We show through analytical and numerical results that this kind of functional diversity across life stages, coupled with continued intra-specific competition, increases the stability of a model community. While functional diversity does not eliminate eventual destabilizing effects of complexity within a community, it provides a structural mechanism that expands system's stability boundary, increasing the number of species a complex community can support. 

\pagebreak
\section*{Model}

\subsection*{\includegraphics[height=4em]{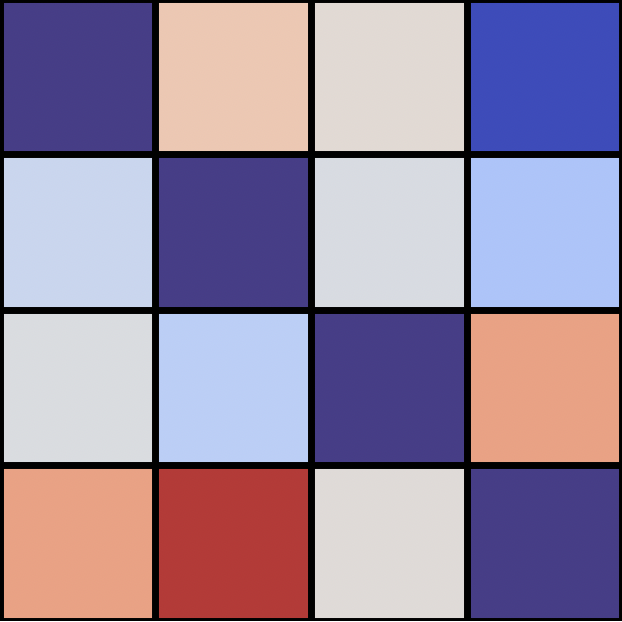} Modeling community dynamics with unstructured populations}

In community ecology, populations of interacting species can be modeled as a system of ordinary differential equations: 
\begin{equation}
\frac{dx_i(t)}{dt} = f_i[\vec{x}(t)]
\end{equation}
where $\vec{x}(t)$ contains the population abundances at time $t$ of each species, and $f_i$ is the function dictating the growth of species $i$ from the abundance of all species present. 

 This system has an equilbrium at $\vec{x}^*$ if all species have zero population change, or  $f_i(\vec{x}^*)=0$ for all $i$. This equilibrium is considered feasible when all populations are nonzero and positive, that is $x_i^* >0$. To determine whether the system will return to this state after a small external perturbation or diverge toward instability, the local asymptotic stability is assessed using the Jacobian matrix $\Jj$ with elements
\begin{equation}
J_{ij}=\frac{\partial f_i[\vec{x}(t)]}{\partial x_j}
\end{equation}
The Jacobian can be evaluated at a feasible equilibrium to reveal the behavior of the system dynamics in in the local neighborhood of $\vec{x}^*$: $\bardown{\Jj}{\vec{x}^*}=\Jj^*$. If the leading eigenvalue of $\Js$ has a strictly negative real part, the system is locally stable and will return to equilibrium. 
Although this mathematical framework results in populations coming to rest with zero fluctuations, ecologically this equilibrium can be understood as the long-term average in a noisy natural environment.

Parameterizing every function $f_i$ and finding $\vec{x}^*$ in a highly diverse natural community is challenging in practice. Instead \citet{may1972} applied random matrix theory to model the interactions of sufficiently large, complex networks of interacting species. When each species is assumed to be a homogeneous population, May's approach is formalized using the generalized Lotka-Volterra (gLV) framework:
\begin{equation}\label{gLV}
    \frac{dx_i(t)}{dt}=x_i(t)\left[r_i+\sum_{j=1}^Sa_{ij}x_j(t)\right]
\end{equation}
where $r_i$ is the intrinsic growth rate of species $i$, and $a_{ij}$ is the per-capita effect of species $j$ on the growth of species $i$. These dynamics are illustrated in \ref{fig:interaction_diagram}(a). The collection of these pairwise inter-species interactions forms an $S\times S$ interaction matrix $\Aa$, where diagonal entries $a_{ii}$ dictate the density-dependent self regulation of each species. Stronger self-regulation (more negative $a_{ii}$) is stabilizing, and local stability cannot be achieved unless a significant portion of species are strongly self-regulating \citep{barabas2017}. This normalization establishes a baseline of intra-specific competition that prevents unbounded population growth in isolation, and effectively isolates the destabilizing impact of the changing inter-specific interactions. Off-diagonal entries (inter-species interactions) are nonzero with probability $C$ and then drawn randomly from a normal distribution with mean $0$ and variance $\sigma^2$; however as long as variables are independent and identically distributed with bounded variance, the spectrum of eigenvalues depends only on the mean and variance of these off-diagonal elements \citep{tao2010}.
\begin{equation}\label{eq:Amatrix}
    \Aa = \begin{bmatrix}
        -1 & a_{12} & a_{13} & \cdots \\
        a_{21} & -1 & a_{23} & \cdots  \\
        a_{31} & a_{32} &  -1 & \cdots \\
        \vdots & \vdots & \vdots & \ddots 
    \end{bmatrix} \qquad a_{ij}=\begin{cases}
        N(0, \sigma^2) & \text{with probability }C\\ 0 & \text{otherwise}
    \end{cases}
\end{equation}

Elements of the Jacobian can be determined from \eqref{gLV} as \begin{align}
    J_{ij} = \frac{\partial}{\partial x_j}\frac{dx_i}{dt} = x_ia_{ij} + \delta_{ij}\left(r_i+x_i\sum_{j} a_{ik}x_k\right)
\end{align}
If an equilibrium exists, it is found at the solution of $f_i(\vec{x}^*)=0$, or 
\begin{equation}\label{eq:onestageconstraint}
0=r_i+\sum_j a_{ij}x_j^*
\end{equation}
By measuring population density $x_{i}$ relative to its equilibrium value, we mathematically translate the system so $\vec{x}^*=1$ represents a feasible equilibrium. While this transformation rescales the distribution of the Jacobian matrix's eigenvalues, it preserves their signs. Local asymptotic stability is strictly dictated by the sign of the real component of the leading eigenvalue, so asymptotic stability is invariant to this rescaling. Changing the measurement scale affects the transient dynamics via the full eigenvalue spectrum, but does not alter the fundamental stability boundary we derive later or stable or unstable outcome of a single community. Although feasibility is also necessary for community persistence \citep{grilli2016}, we assume feasible equilibria to explicitly isolate and explore a system's asymptotic response to external perturbations.

\begin{figure}[!htb]
\includegraphics[width=1\textwidth]{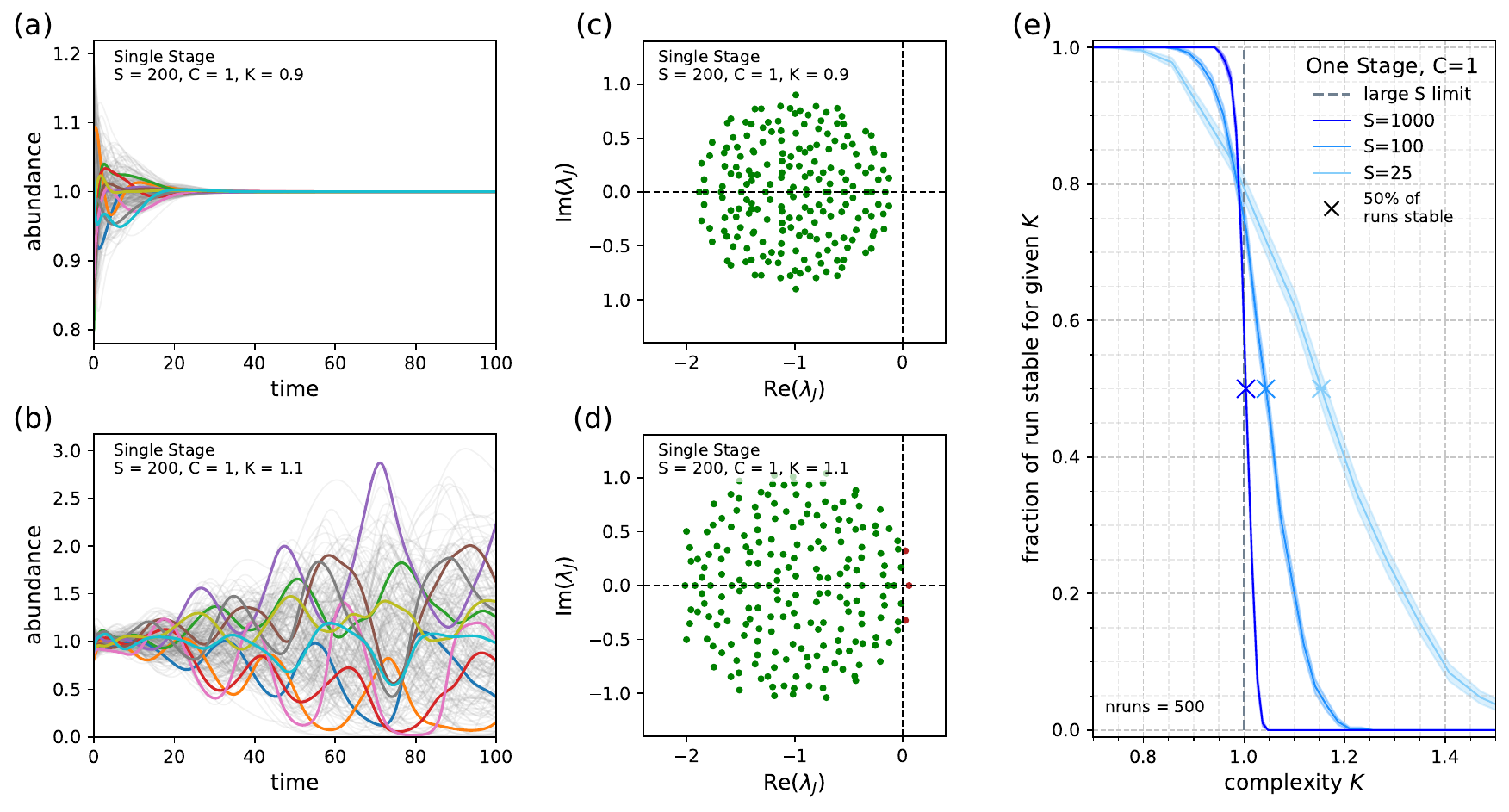}
\caption{\label{fig:scurve_onestage} 
Dynamics and $\Js$ eigenvalues for a single-stage system of 200 species with the same initial conditions in a stable and unstable configuration, determined by the interaction matrix $\Aa$.  The dynamics of 10 species are colored in (a) and (b). Although a realistic community is not expected to settle and stay in a fixed point as shown in (a), this can be understood as a long-term average of stochastic fluctuations. For the system with $K<1$ (a), an initial perturbation returns to equilibrium, and all eigenvalues have negative real component (c). For $K>1$ (b), the system is unstable and several species go extinct. This behavior is predicted by the positive eigenvalues of the equilibrium Jacobian (d). The fraction of random configurations that remain stable for different number of species $S$ (e) are found numerically across a range of $K$. At large $S$, the fraction of stable configurations approaches a sharp transition at $K=1$, confirming previous results of \citet{may1972}.
}
\end{figure}

At this rescaling, at equilibrium the Jacobian is simply \begin{align}
    \Js=\Aa
\end{align} and the local stability is completely determined by the species interactions. From random matrix theory, as $S\to\infty$, the eigenvalues of off-diagonal interactions in $\Aa$ are distributed in a circle in the complex plane with radius equal to the system complexity $K=\sqrt{\sigma^2 SC}$ \citep{girko1985, tao2010}. The diagonal self-regulation terms shift the entire spectrum to be centered at $-1$, so the system becomes unstable when $K$ exceeds 1. This establishes May's famous result: at large $S$, the system is almost surely unstable for $K>1$, and the classical  stability criterion remains: \begin{equation}\label{eq:1stagestabilitycriteria}
    K=\sqrt{\sigma^2SC}< 1
\end{equation} \citep{may1972, Allesina_2012}. The likelihood of stability decreases with increasing system size and interaction strengths, as verified numerically in Figure \ref{fig:scurve_onestage}(e). More diverse systems are thus inherently less stable.
For systems with finite $S$, the position of the leading eigenvalue fluctuates according to the Tracy-Widom distribution \citep{tracy1994}. These finite-size fluctuations scale with $S^{-\frac{2}{3}}$, so the abrupt transition at large $S$ is smoothed for finite community size \citep{majumdar2014}, as seen in Figure \ref{fig:scurve_onestage}(e). Finite systems thus have nonzero probability of stability at and above the theoretical stability limit, but follow the same trend of decreased stability for increased intraction variance $\sigma$.

\subsection*{Adding stage structure}

\begin{figure}[!htb]
\includegraphics[width=0.99\textwidth]{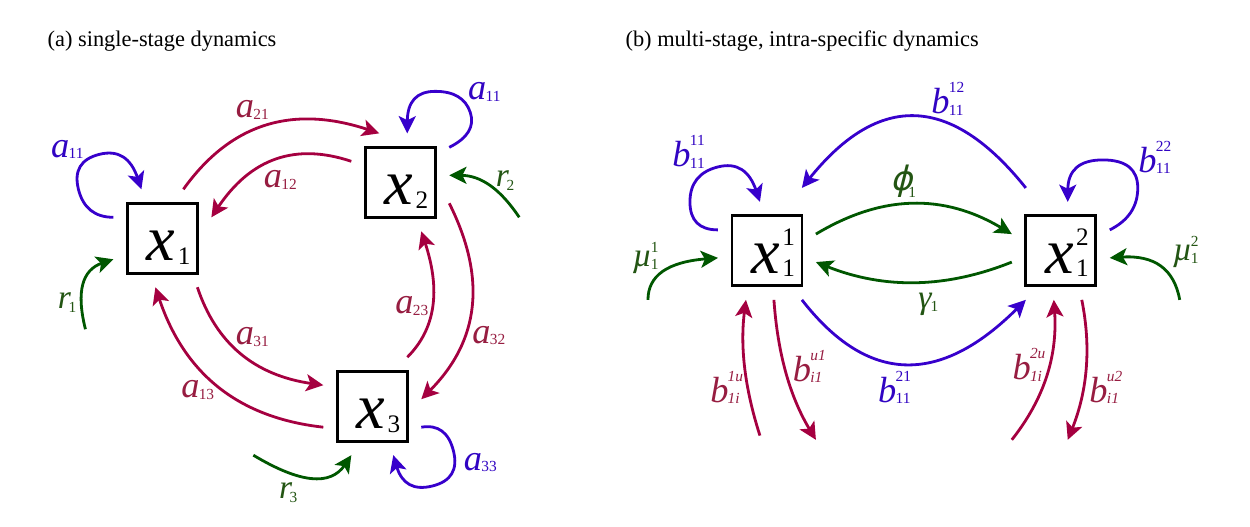}
\caption{\label{fig:interaction_diagram} 
Interactions between (red) and within (blue) species in the single- and multi-stage cases. (a) When each species is modeled as a single life stage, there is one set of interactions between two species, $a_{ij}$ and $a_{ji}$. All demographic processes, including birth and death, are aggregated into a single intrinsic growth rate $r_i$ (green). (b) When multiple stages are introduced, both inter-specific (red) and intra-specific (blue) interactions are now explicitly dependent on the stages involved.  Intra-species dynamics are explicitly regulated via a life cycle that governs transitions between stages (green).
}

\end{figure}

Now we consider that each species is not a homogeneous population, but is composed of $L$ life stages. The population density of each stage $\alpha\in 1:L$ of species $i$ is $x_i^\alpha.$ The complete population vector is \begin{equation}
    \vec{x} = \left(x_1^1, x_1^2, \dots, x_1^L, x_2^1, x_2^2, \dots, x_2^L, \dots, x_S^L\right)
\end{equation}  
Each stage has some regulation rate $\mu_i^\alpha <0$, arising from mortality. Younger stages $\alpha\leq L-1$ grow into the next stage at a maturation rate $\gamma^\alpha_i>0$. In older stages $\alpha>0$ in which individuals can reproduce, they contribute to the juvenile population with a fecundity rate $\phi^u_i>0$. These intra-specific, inter-stage dynamics are illustrated in \ref{fig:interaction_diagram}(b). The growth rate vector $\vec{r}$ of the single-stage case is now replaced by a block-diagonal demographic matrix $\Rr$, to capture intra-specific linear dynamics. In the case where each species has 2 stages, we have 
\begin{equation}
\Rr=\begin{bmatrix}
\mu_1^1 & \phi_1 & \\ \gamma_1 & \mu_1^2 & \\
& & \mu_2^1 & \phi_2 \\ & & \gamma_2 & \mu_2^2 & \\ & & & & \ddots
\end{bmatrix}
\end{equation} where outside the block-diagonal, all entries are always zero.

We also construct a new multi-stage interaction matrix $\Bb$ to account for the stage structure with dimensions $LS\times LS$. In the two stage case,
\begin{equation}
\Bb=\begin{bmatrix} 
-1 & b_{11}^{12} & b_{12}^{11} & b_{12}^{12} & \cdots \\
b_{11}^{21} & -1 & b_{12}^{21} & b_{12}^{22} & \cdots \\
b_{12}^{11} & b_{12}^{12} & -1 & b_{22}^{12} & \cdots \\
b_{12}^{21} & b_{12}^{22} & b_{22}^{21} & -1 & \cdots \\
\vdots & \vdots & \vdots & \vdots &\ddots
\end{bmatrix}
\end{equation}
where $b_{ij}^{\alpha\beta}$ describes the effect of stage $\beta$ of species $j$ on the population density of stage $\alpha$ of species $i$. The specific entries of the interaction matrix $\Bb$ are dependent on assumptions of how interactions vary across life stages, and is expanded upon further in following sections. Connectance is set to $C=1$ for all cases moving forward. Cases with $C\neq 1$ are explored further in the Supplementary Information, and a suitible generalization of our results still applies. We maintain the same self-regulation for same-species, same-stage interactions as in the single-stage case, setting diagonal entries as $b_{ii}^{\alpha\alpha}=-1$. 

For the following, we focus on a community of species each with two stages. In this case, stage 1 of species $i$ grows into stage 2 with rate $\gamma_i$, and stage 2 reproduces individuals of stage 1 with rage $\phi_i$. The system dynamics then follow \begin{align}
\frac{dx^{1}_{i}}{dt}=\mu_i^1x_i^1 +\phi_i x_i^2+x_{i}^1\left(b_{i1}^{11}x_1^1 + b_{i1}^{12}x_1^2 + b_{i2}^{11}x_2^1+ b_{i2}^{12}x_2^2+\dots\right) \qquad \text{for stage 1 of species $i$} \\
\frac{dx^{2}_{i}}{dt}=\gamma_i x_i^1+\mu_i^2x_i^2 +x_{i}^2\left(b_{i1}^{21}x_1^1 + b_{i1}^{22}x_1^2 + b_{i2}^{21}x_2^1+ b_{i2}^{22}x_2^2+\dots\right) \qquad \text{for stage 2 of species $i$}
\end{align}
 In general, a system of $S$ species each with $L$ stages is governed by
\begin{equation}\label{eq:2stagedtnamics}
\frac{dx^{\alpha}_{i}}{dt}=\sum_{j,\beta}R_{ij}^{\alpha\beta}x_{j}^\beta+x_{i}^u\sum_{j,\beta} b_{ij}^{\alpha\beta}x_{j}^\beta
\end{equation}
Where the summation is taken over every stage $(\beta=1:L)$ of every species $(j=1:S)$ to capture all $S\cdot L$ interactions that affect stage $\alpha$ of species $i$.

We again assume a feasible equilibrium point, and manipulate $\Rr$ in order to focus on the stability at this point. For an equilibrium to exist at $\vec{x}^*=1$, we must have \begin{align}\label{eq:2stagecontraint}
    \sum_{j,\beta}b_{ij}^{\alpha\beta} &= -\sum_{j,\beta}R_{ij}^{\alpha\beta} 
\end{align} 
found by setting \eqref{eq:2stagedtnamics} equal to $0$ when $x_i^\alpha=1$ for each stage of each species.
This results in the constraints:  \begin{align}
    \sum_{j,\beta}b_{ij}^{1\beta} &=  -\mu_i^1-\phi_i \\
    \sum_{j,\beta}b_{ij}^{2\beta} &= -\gamma_i-\mu_i^2 
\end{align}
Further, we require that these quantities are biologically realistic, so $\mu_i^1, \mu_i^2 \leq 0$ and $\phi_i, \gamma_i \geq 0$. Values in $\Rr$ are still under-constrained, so we set the fecundity and growth rates as \begin{align}\label{eq:Rvalmins}
    \phi_i=\text{max}\left(0, -\sum_{j,\beta}b_{ij}^{1\beta}\right)+\Delta \qquad  \gamma_i=\text{max}\left(0, -\sum_{j,\beta}b_{ij}^{2\beta}\right)+\Delta
\end{align} for some $\Delta\geq 0$, and then find the regulation rates as \begin{align}
    \mu_i^1=-\phi_i-\sum_{j,\beta}b_{ij}^{1\beta} \qquad \mu_i^2=-\phi_i-\sum_{j,\beta}b_{ij}^{2\beta}
\end{align} which guarantees a feasible equilibrium at $\vec{x}^*=1$. 
The value of $\Delta$ sets the minimum magnitude of any rate in $\Rr$, and thus $\Delta$ sets the timescale of the life cycles of species in the community.

Elements of the Jacobian of the two-stage system can be determined from \eqref{eq:2stagedtnamics} as \begin{equation}
    J_{ij}^{\alpha\beta} =\frac{\partial}{\partial x_j^\beta}\frac{dx^{\alpha}_{i}}{dt}
    = R^{\alpha\beta}_{ij}+x^{\alpha}_iB^{\alpha\beta}_{ij}+\delta_{ij}\delta_{\alpha\beta}\sum_{k,\xi}b_{ik}^{\alpha\xi}
\end{equation}
At $\vec{x}^*=1$, with the constraint of \eqref{eq:2stagecontraint}, the equilibrium Jacobian is \begin{align}\label{eq:2stagejac}
    \Js &= \Rr + \Bb + \text{diag}(\Bb \cdot \vec{1})
\end{align}

where $\vec{1}$ is a vector of $1$, of length $LS$. The eigenvalues of $\Js$ can now be evaluated to determine the linear stability of the system near the feasible equilibrium.

\subsubsection*{\includegraphics[height=4em]{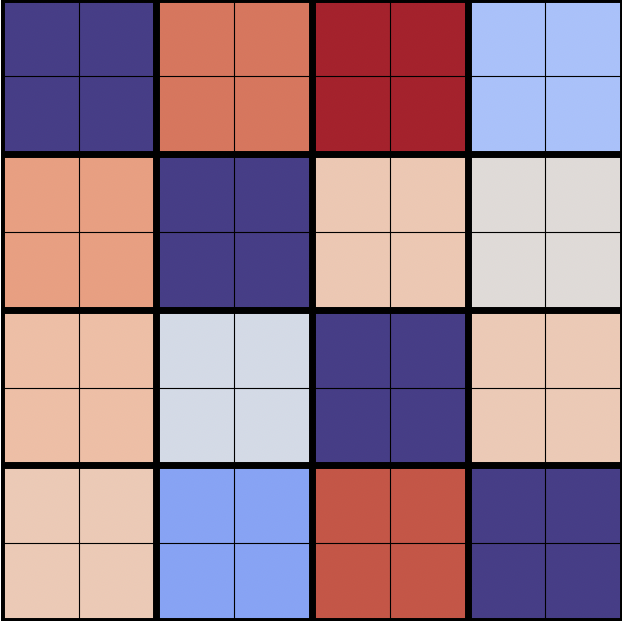} Identically interacting stages}

The simplest multi-stage model used assumes that inter-species interactions are independent of life stage. Then $b_{\alpha\beta}^{uv}=b_{ij}$: two species interact the same regardless of the specific life stages involved. Self-regulation then also extends to intra-specific, inter-stage interactions ($b_{ii}^{\alpha\beta}=-1$). The full interaction matrix $\Bb_1$ is
\begin{equation}\label{eq:B1}
\Bb_1=\begin{bmatrix} 
-1 & -1 & b_{12} & b_{12} & \cdots \\
-1 & -1 & b_{12} & b_{12} & \cdots \\
b_{21} & b_{21} & -1 & -1 & \cdots \\
b_{21} & b_{21} & -1 & -1 & \cdots \\
\vdots & \vdots & \vdots & \vdots &\ddots
\end{bmatrix} \qquad b_{ij}=
        N(0, \sigma^2)\text{ for } i\neq j
\end{equation}
with exactly repeated rows and columns for the different life stages. Therefore, the distinction between life stages only affects the demographic regulation of a species via the matrix $\Rr$. 

\subsubsection*{\includegraphics[height=4em]{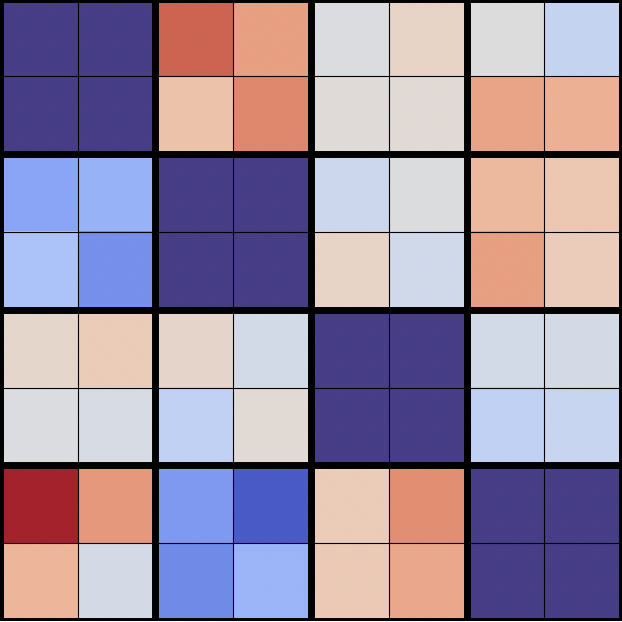} Stage-dependent interactions}

Here we relax the strong assumption that inter-species interactions are identical across stages, opening  up a range of new interaction matrices. When a species' ability to compete, cooperate, or avoid predation changes across its life stages, inter-specific interaction terms will vary as well.  If there is little difference in how  species $i$ is impacted by the populations of other species during its life, $b_{ij}^{\alpha\beta}$ will have little or no variation across stages $\alpha$. But when the effect of species $j$ changes throughout the life cycle of $i$, interaction terms may be different, completely independent, or even qualitatively opposite for different stages.

To generalize this in our model, the $L\times L$ stage-dependent effects of one species on another are correlated across stages by a factor $\rho$, such that the interactions between species $i$ and $j$ have a covariance of $\text{Cov}(b_{ij}^{\alpha\beta}, b_{ij}^{\alpha'\beta'})=\rho\sigma^2$ across the $L^2$ terms. In the two-stage system, the interaction matrix $\Bb_\rho$ is 
\begin{equation}\label{eq:Brhomatrix}
\Bb_{\rho}=\begin{bmatrix} 
-1 & -1 & b_{12}^{11} & b_{12}^{12} & \cdots \\
-1 & -1 & b_{12}^{21} & b_{12}^{22} & \cdots \\
b_{21}^{11} & b_{21}^{12} & -1 & -1 & \cdots \\
b_{21}^{21} & b_{21}^{22} & -1 & -1 & \cdots \\
\vdots & \vdots & \vdots & \vdots &\ddots
\end{bmatrix}\qquad \text{Cov}(b_{ij}^{\alpha\beta}, b_{ij}^{\alpha'\beta'})=\rho\sigma^2
\end{equation} 
When $\rho=1$, $b_{ij}^{\alpha\beta}=b_{ij}^{\alpha'\beta'}$, returning the system to the case of identically interacting stages. At $\rho=0$, $\text{Cov}(b_{ij}^{\alpha\beta}, b_{ij}^{\alpha'\beta'})=0$ and the interactions are completely independent, as if different stages interact as entirely different species. A higher value of $\rho$ indicates that different life stages behave more similarly. Mathematically, $\rho$ is maximally bounded at $1$, for perfect correlation, and minimally bounded by $\frac{-1}{L^2-1}$, where stages interact as differently as possible. We use an equi-correlation structure that applies the covariance uniformly across the network as a mean-field approximation.

This correlation $\rho$ is used to tune the stage-dependent heterogeneity in inter-specific interactions, which can be thought of as the functional diversity across life stages of a species. When $\rho$ is decreased, different stages of the same species interact more differently, or even have opposite interaction signs for interactions with another species.

\subsubsection*{\includegraphics[height=4em]{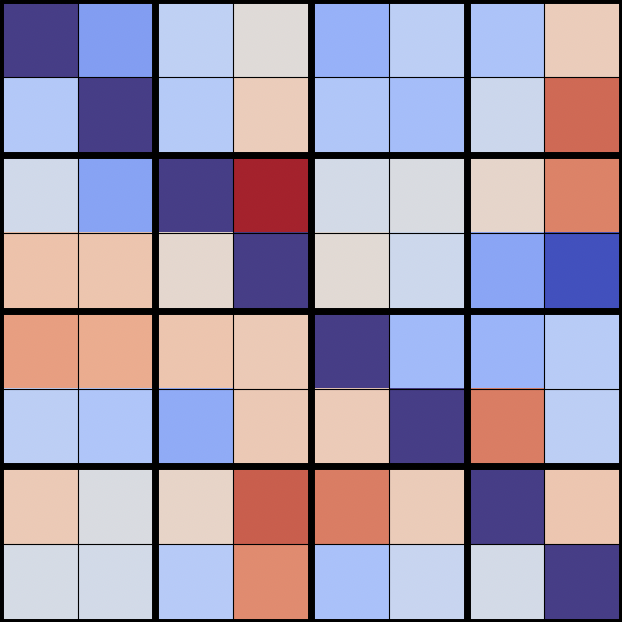} Random intra-specific interactions}

Finally, we consider cases when same-species interactions are not fixed across life stages. Instead of fixing inter-stage interactions to be strongly negative, we instead now draw these interactions randomly, such that $b_{ii}^{\alpha\beta}\sim N(0, \sigma^2)$ for $\alpha\neq \beta$. Under this condition, distinct stages of the same species interact with each other as if they were entirely distinct species. Although we may expect intra-species competition among all stages, perhaps when the inter-stage correlation is small stages may be so different that they do not necessarily compete with each other strongly. The only mathematical distinction between this structured model and a classical one-stage system with $2S$ species now is the presence of the demographic matrix $\Rr$, which dictates the life cycle between stages. Taking all inter- and intra-specific interactions across stages as completely independent, the $\Bind$ is 
\begin{equation} \label{eq:Bindmatrix}
\Bind=\begin{bmatrix} 
-1 & b_{11}^{12} & b_{12}^{11} & b_{12}^{12} & \cdots \\
b_{11}^{21} & -1 & b_{12}^{21} & b_{12}^{22} & \cdots \\
b_{12}^{11} & b_{12}^{12} & -1 & b_{22}^{12} & \cdots \\
b_{12}^{21} & b_{12}^{22} & b_{22}^{21} & -1 & \cdots \\
\vdots & \vdots & \vdots & \vdots &\ddots
\end{bmatrix} \qquad b_{ij}^{\alpha\beta} \sim N(0,\sigma^2)
\end{equation}

\subsubsection*{Model summary}

We now have four cases to compare the effects of stage structure in a generalized Lotka-Volterra system, summarized in Table \ref{tab:model}. The extensively studied one-stage case (with interaction matrix $\Aa$) provides a baseline with well-understood stability criteria. We first generalize this to a multi-stage model where stages interact identically with other species ($\Bb_1$), but are demographically linked via the growth and fecundity rates of each stage. We then relax the constraint on inter-species interactions, and allow the stage-dependent effects of one species on another to vary, correlated by a factor of $\rho$ ($\Bb_{\rho}$). Finally, we consider the extreme case in which inter- and intra-species interactions are completely uncoupled across stages, effectively treating different life stages as distinct species ($\Bind$). 

\begin{table}[h!]
    \caption{Model types used for interaction matrix}
    \label{tab:model}
\begin{center}
\begin{tabular}{ |>{\centering\arraybackslash}m{2.5cm} | >{\centering\arraybackslash}m{3cm}|m{9cm}|}
\hline
    \textbf{Visual} &\textbf{Model}& \multicolumn{1}{c|}{\textbf{Description}} \\ 
 \hline
 \adjustbox{margin=2mm}{\includegraphics[width=2.2cm]{figs/icons/onestage_black.png}}
 & Single stage ($\Aa$) & Each species treated as a homogeneous population with a fixed growth rate and an $S\times S$ interaction matrix. \\  
 \hline
 \adjustbox{margin=2mm}{\includegraphics[width=2.2cm]{figs/icons/rho_1_black.png}}
 & Identically interacting stages ($\Bb_1$) & Multiple life stages are differentiated by their growth, death, fecundity. Inter-species interactions are independent of life stage, and only depend on the two species interacting. \\  
 \hline
 \adjustbox{margin=2mm}{\includegraphics[width=2.2cm]{figs/icons/rho_08_black.png}}
 & Stage-dependent interactions ($\Bb_{\rho}$) & Multiple life stages are differentiated by their growth, death, fecundity. Interactions depend on the stages interacting, and the different stage-dependent interactions between two species are correlated by a factor of $\rho$.  \\  
 \hline
 \adjustbox{margin=2mm}{\includegraphics[width=2.2cm]{figs/icons/ind_black.png}}
 & Random intra-specific interactions ($\Bind$) &  Multiple life stages are differentiated by their growth, death, fecundity. Different stages interact as completely distinct species. This interaction matrix is mathematically equivalent to  the stage-stage system with $2S$ species.
 \\  
 \hline
\end{tabular}
\end{center}
\end{table}

\pagebreak
\section*{Results}

Following the analysis used to obtain the one-stage result \eqref{eq:1stagestabilitycriteria}, a combination of analytical and numerical methods are used to analyze our different cases of two-stage systems.

\subsection*{\includegraphics[height=4em]{figs/icons/rho_1_black.png} Identically interacting stages: $\rho=1$}

We first consider that life stages only affect intra-specific dynamics via the demographic matrix $\Rr$, and different stages of a species behave identically toward other species \eqref{eq:B1}. 
The only mathematical differentiation from the classical one-stage gLV model is the inclusion of inter-stage terms in $\Rr$. At the feasible equilibrium of $\vec{x}^*=\vec{1}$, from \eqref{eq:2stagejac} the Jacobian is now 
\begin{align}\label{Js_p1}
    \Js_1 &= \Rr + \Bb_1 + \diag{\Bb_1\cdot\vec{1}}\\
    \Js_1 &= \begin{bmatrix}
        -\phi_1-1 & \phi_1-1  & b_{12} & b_{12} &\cdots\\ \gamma_1-1 & -\gamma_1-1  & b_{12} & b_{12} & \cdots \\b_{21}&b_{21} & -\phi_2-1 & \phi_2-1 & \cdots\\ b_{21} & b_{21} & \gamma_2-1  & -\gamma_2^2-1  & \cdots\\ \vdots& \vdots& \vdots& \vdots& \ddots
    \end{bmatrix}
\end{align}
The eigenvalues determining local linear stability come from solving the characteristic equation:
\begin{align}
    \det(\Js-\lambda\Id)=0
\end{align} 
and partition cleanly into two sets: those of matrix $\Bb$ and those of the matrix $\Rr'=\Rr+\diag{\Bb_1\cdot\vec{1}}$.

\begin{align}\label{eq:eigsrho1}
    \{\lambda_{J^*} \}&= \begin{cases}
        \{\lambda_B\}=2\{\lambda_A\} \\
        \{\lambda_{R'}\} = -\phi_i-\gamma_i, \quad  i=1:S
    \end{cases}
\end{align} 
Where $\{\lambda_A\}$ is the set of eigenvalues of the equivalent one-stage $S\times S$ interaction matrix with $a_{ij}=b_{ij}$ (for the detailed derivation, see Supplementary Information). Thus eigenvalues $\{\lambda_B\}$ are distributed in a circle centered at $-2$ with radius $2\sigma\sqrt{SC}=2K$. The symmetry of identical interactions across stages doubles the effective interaction of each species (doubling the radius of distribution), and also doubles the intra-specific competition term (shifting the center to $-2$).

The eigenvalues $\{\lambda_{R'}\}$ depend only on demographic rates and will always be real and negative, as $\phi_i, \gamma_i >0$ for all species. This is a mathematical consequence of enforcement of the feasibility condition. While highly negative eigenvalues accelerate a system's return to equilibrium following a perturbation, they cannot reverse the destabilizing effect of a positive leading eigenvalue \citep{may1973}. Consequently, the asymptotic stability of the two-stage system  with $\rho=1$ depends only on the eigenvalues of the interaction matrix $\Aa$ from the analogous one-stage case. 

For identically interacting stages, the classical stability criterion remains unchanged, and in the large $S$ limit the system is locally stable around $\vec{x}^*=\vec{1}$ for $K=\sigma\sqrt{SC}<1$ \eqref{eq:1stagestabilitycriteria}. This is confirmed numerically in Figure \ref{fig:Scurves}(a) This result illustrates an important finding: simply partitioning a population into demographic compartments ($\rho=1$) does not inherently buffer a community against the destabilizing effects of complexity.

\subsection*{\includegraphics[height=4em]{figs/icons/rho_08_black.png} Varying} interactions across stages: $\rho\neq 1$

While perfect symmetry ($\rho=1$) provides a useful mathematical limit, most real-world populations exhibit some degree of  differentiation between life stages \citep{zhao2014}. By varying $\rho$, we can capture this variable functional diversity and model stage-dependent interactions that are partially correlated ($0<\rho<1$), uncorrelated ($\rho=0$), or even anti-correlated ($\rho<0$) across a species' life cycle.

Inter-species interactions are still drawn from $N(0, \sigma^2)$, but now maintain a covariance of $\cov{b_{ij}^{\alpha\beta},b_{ij}^{\alpha'\beta'}}=\rho\sigma^2$ for the set of interaction effects of species $j$ on species $i$ \eqref{eq:Brhomatrix}.

For $B_{rho}$ with $\rho\neq 1$, the Jacobian from $\eqref{eq:2stagejac}$ is \begin{align}
    \Js_{\rho} &= \Rr + \Bb_\rho + \diag{\Bb_\rho\cdot\vec{1}} \\
    \Js_{\rho} &= \begin{bmatrix}
        -1-\phi_1& -1+\phi_1 & b_{12}^{11} & b_{12}^{12} & \cdots \\
-1+\gamma_1 & -1-\gamma_1 & b_{12}^{21} & b_{12}^{22} & \cdots \\
b_{12}^{11} & b_{12}^{12} & -1-\phi_2 & -1+\phi_2 & \cdots \\
b_{12}^{21} & b_{12}^{22} & -1+\gamma_2 & -1-\gamma_2 & \cdots \\
\vdots & \vdots & \vdots & \vdots &\ddots
    \end{bmatrix}
\end{align}
And the eigenvalues come from the characteristic equation $\det({\Js_\rho}-\lambda \Id)=0$, where \begin{align}
    \det(\Js_\rho-\lambda \Id)&=\begin{vmatrix}
        -1-\phi_1-\lambda & -1+\phi_1 & b_{12}^{11} & b_{12}^{12} & \cdots \\
-1+\gamma_1 & -1-\gamma_1-\lambda & b_{12}^{21} & b_{12}^{22} & \cdots \\
b_{12}^{11} & b_{12}^{12} & -1-\phi_2-\lambda & -1+\phi_2 & \cdots \\
b_{12}^{21} & b_{12}^{22} & -1+\gamma_2 & -1-\gamma_2-\lambda & \cdots \\
\vdots & \vdots & \vdots & \vdots &\ddots
    \end{vmatrix} 
\end{align} After some elementary row operations which preserve the determinant (detailed in Supplementary Information), we have 
\begin{align}
    \det(\Js_\rho-\lambda \Id) &= \begin{vmatrix}
       -2-\lambda & b_{12}^{11}+b_{12}^{12} & \cdots & -1+\phi_1 & b_{12}^{12} & \cdots \\ b_{21}^{11} + b_{21}^{12} & -2-\lambda & \cdots & b_{21}^{12} & -1+\phi_2 & \cdots \\ 
       \vdots & \vdots & \ddots & \vdots & \vdots &\ddots \\
       0 & b_{12}^{21}+b_{12}^{22}-b_{12}^{11}-b_{12}^{12} & \cdots & -\gamma_1-\phi_1-\lambda & b_{12}^{22}-b_{12}^{12} & \cdots \\ b_{21}^{21}+b_{21}^{22}-b_{21}^{11}-b_{21}^{12} & 0 & \cdots & b_{21}^{22}-b_{21}^{12} & -\gamma_2-\phi_2-\lambda & \cdots \\
\vdots & \vdots & \vdots & \vdots&  \vdots &\ddots
    \end{vmatrix} 
\end{align} 

If we set the magnitude of each parameter in $\Rr$ to its minimum ($\gamma_i, \phi_i \geq \Delta$, from \eqref{eq:Rvalmins}), then \begin{align}
    \det(\Js_\rho-\lambda \Id) &= \begin{vmatrix}
       -2-\lambda & b_{12}^{11}+b_{12}^{12} & \cdots & -1+\Delta & b_{12}^{12} & \cdots \\ b_{21}^{11} + b_{21}^{12} & -2-\lambda & \cdots & b_{21}^{12} & -1+
       \Delta & \cdots \\ 
       \vdots & \vdots & \ddots & \vdots & \vdots &\ddots \\
       0 & b_{12}^{21}+b_{12}^{22}-b_{12}^{11}-b_{12}^{12} & \cdots & -2\Delta-\lambda & b_{12}^{22}-b_{12}^{12} & \cdots \\ b_{21}^{21}+b_{21}^{22}-b_{21}^{11}-b_{21}^{12} & 0 & \cdots & b_{21}^{22}-b_{21}^{12} & -2\Delta-\lambda & \cdots \\
\vdots & \vdots & \vdots & \vdots&  \vdots &\ddots
    \end{vmatrix} 
\end{align}

Here we use the formula for the determinant of a partitioned matrix , where  \begin{align} \label{eq:blockdet}
    \text{for} \quad \mathbf{M}=\begin{bmatrix}
        \mathbf{W} & \mathbf{X} \\ \mathbf{Y} & \mathbf{Z} 
    \end{bmatrix},\qquad  \det(\mathbf{M})=\begin{cases}
        \det(\mathbf{W})\det(\mathbf{Z-YW^{-1}X}) & \text{if } \det{\mathbf{W}\neq 0} \\
        \det(\mathbf{Z})\det(\mathbf{W-XZ^{-1}Y})& \text{if } \det{\mathbf{Z}\neq 0} \\
    \end{cases}
\end{align} 
\citep{Abadir2005}.
To isolate the effects of the stage-dependent interactions, we take the limit of $\Delta\to\infty$ for our analytical results. Large $\Delta$ gives fast demographic rates ($\phi_i, \gamma_i$) relative to inter-specific interactions. 
In this limit, our characteristic equation is solved by  
\begin{align}
    &0 = \det(\mathbf{Z-YW^{-1}X}) \label{eq:Zrho}
    \\ \text{or }\ \ &0 = \det(\mathbf{W-XZ^{-1}Y})\label{eq:Wrho}
\end{align}where \begin{align}
     \mathbf{Z-YW^{-1}X}&= \begin{bmatrix}
     -2\Delta-\lambda & \frac{1}{2}\left(3b_{12}^{22}-3b_{12}^{12}+b_{12}^{22}-b_{12}^{11}\right) & \cdots \\ 
        \frac{1}{2}\left(3b_{21}^{22}-3b_{21}^{12}+b_{21}^{22}-b_{21}^{11}\right) & -2\Delta-\lambda & \cdots \\ \vdots & \vdots & \ddots 
     \end{bmatrix}\\ \nonumber\\ 
     \mathbf{W-XZ^{-1}Y}&=
     \begin{bmatrix}
         -2-\lambda & \frac{1}{2}\left(b_{12}^{11}+b_{12}^{12}+b_{12}^{21}+b_{12}^{22}\right) & \cdots \\
         \frac{1}{2}\left(b_{21}^{11}+b_{21}^{12}+b_{21}^{21}+b_{21}^{22}\right) & -2-\lambda & \cdots \\ \vdots & \vdots & \ddots
     \end{bmatrix}
\end{align}

Each solution gives a set of $S$ eigenvalues distributed in the complex plane, with $S$ near $-2\Delta$ from \eqref{eq:Zrho}, and $S$ centered at $-2$ with radius $\sigma\sqrt{SC \cdot(1+3\rho)}$ from \eqref{eq:Wrho}. In the limit of large $S$ and $\Delta$, the leading eigenvalue will be negative if $\sigma\sqrt{SC \cdot (1+3\rho)}<2$, giving the new stability threshold \begin{align}\label{eq:largedeltaresult}
    K<\frac{2}{\sqrt{1+3\rho}}
\end{align} Where again $K=\sigma\sqrt{SC}$ is the complexity to compare with the single-stage case. Notably, off-diagonal elements of $\mathbf{W-XZ^{-1}Y}$ determine the critical radius here and are a function of the sum of interaction terms of one species on another, implying the importance of the average interaction of one species on another. Coupled with the effective intra-specific competition of $-2$, this gives the increased stability threshold. As $\rho$ decreases toward 0 and into negative values for increasingly uncorrelated or anti-correlated stages, the overall spectral radius of the Jacobian contracts. At the limit of $\Delta\to\infty$ for $\rho=-\frac{1}{3}$, \eqref{eq:largedeltaresult} diverges and any configuration of $\Bb_{-\frac{1}{3}}$ is predicted to be stable as the spectral radius approaches 0. For further details, see  Supplementary Information.

We can understand this result using a new \textit{effective} interaction variance of $(\sigma')^2=\sigma^2(1+3\rho)$. When life stages are functionally different, the decreased covariance reduces this aggregated interaction variance, because the ontogenetic niche shifts diffuse interaction strengths across a life cycle.

For finite $\Delta$, results are found numerically, shown in Figure \ref{fig:deltafig}. For $\rho=1$, the stability threshold is independent of $\Delta$, matching \eqref{eq:eigsrho1}. Large $\Delta$ creates some timescale separation, such that demographic changes happen much faster than interactions. As $\rho$ decreases, larger $\Delta$ is required to use \eqref{eq:largedeltaresult} as a good approximation of the stability-complexity relationship of the system. But even at small $\Delta\sim 1$ the qualitative result of increased stability threshold for decreased $\rho$ holds, and lower stage-dependent interaction correlation amplifies the stabilizing effect of the stage structure (Figure \ref{fig:deltafig}(b)).

\subsection*{\includegraphics[height=4em]{figs/icons/ind_black.png} Random Intra-specific interactions}

Now we consider the case in which both inter- and intra-specific interactions are entirely uncorrelated ($\rho=0$), so individuals interact with other life stages of their own species as they would an entirely different species ($b_{ij}^{\alpha\beta}\sim N(0, \sigma^2)$ for $i\neq j$ or $\alpha\neq \beta$). Under these conditions, the interaction matrix $B_{ind}$ is mathematically equivalent to the one-stage interaction matrix for a system of $2S$ species.
Different stages of the same species interact as hetero-specific individuals, and are only linked by the demographic processes of growth and fecundity in $\Rr$.

Using the interaction matrix in \eqref{eq:Bindmatrix} we again evaluate the Jacobian at equilibrium:
\begin{align}\label{eq:jac_ind}
    \Js_{ind} &= \Rr + \Bb_{ind} + \diag{\Bb_{ind}\cdot\vec{1}}\\
    \Js_{ind} &= \begin{bmatrix}
        -\phi_1-1 & \phi_1+b_{11}^{12} &b_{12}^{11}&b_{12}^{12} & \cdots\\ \gamma_1+b_{11}^{21} & -\gamma_1-1 & b_{12}^{21}&b_{12}^{22} & \cdots \\ 
        b_{21}^{11} & b_{21}^{12} &-\phi_2-1&\phi_2+b_{22}^{12} & \cdots\\ b_{21}^{21} & b_{21}^{22} & \gamma_1+b_{22}^{21}&-\gamma_2-1 & \cdots \\ 
        \vdots & \vdots & \vdots & \vdots & \ddots 
    \end{bmatrix}
\end{align}
And following our previous procedure, we look at the characteristic equation $\det(\Js_{ind}-\lambda \Id)=0$ to find the eigenvalues.
\begin{align}
    \det(\Js_{ind}-\lambda \Id) &=\begin{vmatrix}
        -\phi_1-1-\lambda & \phi_1+b_{11}^{12} &b_{12}^{11}&b_{12}^{12} & \cdots\\ \gamma_1+b_{11}^{21} & -\gamma_1-1-\lambda & b_{12}^{21}&b_{12}^{22} & \cdots \\ 
        b_{21}^{11} & b_{21}^{12} &-\phi_2-1-\lambda&\phi_2+b_{22}^{12} & \cdots\\ b_{21}^{21} & b_{21}^{22} & \gamma_1+b_{22}^{21}&-\gamma_2-1-\lambda & \cdots \\ 
        \vdots & \vdots & \vdots & \vdots & \ddots 
    \end{vmatrix} 
\end{align}
Using elementary row operations and taking each parameter in $\Rr$ to its minimum of $\Delta$ as before, we have 
\begin{align}
    \det(\Js_{ind}-\lambda \Id) &= \begin{vmatrix}
        -1+b_{11}^{12}-\lambda & b_{12}^{11}+b_{12}^{12} & \cdots & b_{11}^{12}+\Delta & b_{12}^{12} & \cdots \\
        b_{21}^{11}+b_{21}^{12} & -1+b_{22}^{12}-\lambda & \cdots &b_{12}^{21} & b_{22}^{21}+\Delta & \cdots \\
        \vdots & \vdots & \ddots & \vdots & \vdots & \ddots \\
        b_{11}^{21}-b_{11}^{12} & b_{12}^{21}+b_{12}^{22}-b_{12}^{11}-b_{12}^{12} & \cdots & -1-b_{11}^{12}-2\Delta-\lambda & b_{12}^{22}-b_{12}^{12} & \cdots \\
        b_{21}^{21}+b_{21}^{22}-b_{21}^{11}-b_{21}^{12} & b_{22}^{21}-b_{22}^{12} & \cdots & b_{21}^{22}-b_{21}^{12} & -1-b_{22}^{12}-2\Delta-\lambda & \cdots \\
        \vdots & \vdots & \ddots & \vdots & \vdots & \ddots \\
    \end{vmatrix}
\end{align}
Again using the formula for the determinant of a partitioned matrix \eqref{eq:blockdet}, we have solutions to the characteristic equations from 
\begin{align}
    &0 = \det(\mathbf{Z-YW^{-1}X})\label{eq:Zind} 
    \\ \text{or }\ \ &0 = \det(\mathbf{W-XZ^{-1}Y})\label{eq:Wind} 
\end{align}where \begin{align}
     \mathbf{Z-YW^{-1}X}&= \begin{bmatrix}
     -1-2\Delta -\frac{1}{2}(b_{11}^{12}+b_{11}^{21})-\lambda & \frac{1}{2}(b_{12}^{11}+b_{12}^{22}-b_{12}^{12}-b_{12}^{21}) & \cdots \\
        \frac{1}{2}(b_{21}^{11}+b_{21}^{22}-b_{21}^{12}-b_{21}^{21}) & -1-2\Delta -\frac{1}{2}(b_{22}^{12}+b_{22}^{21})-\lambda & \cdots \\ \vdots & \vdots & \ddots
     \end{bmatrix}\\ \nonumber\\ 
     \mathbf{W-XZ^{-1}Y}&=
    \begin{bmatrix}
        -1+\frac{1}{2}(b_{11}^{21}-b_{11}^{12})-\lambda & \frac{1}{2}\left(b_{12}^{21}+b_{12}^{22}+b_{12}^{11}+b_{12}^{12}\right) & \cdots \\
        \frac{1}{2}\left(b_{21}^{21}+b_{21}^{22}+b_{21}^{11}+b_{21} ^{12}\right) & -1+\frac{1}{2}(b_{22}^{21}-b_{2}^{12})-\lambda & \cdots \\ \vdots & \vdots & \ddots 
    \end{bmatrix}
\end{align}

For large $S$ and $\Delta$, solutions to the characteristic equation satisfying \eqref{eq:Zind} give eigenvalues near $-2\Delta$ with finite radius. These will always be negative in the limit of large $\Delta$. Off-diagonal elements of $\mathbf{W-XZ^{-1}Y}$ come from the normal distribution $N(0, \sigma^2)$ resulting from the factor of $\frac{1}{2}$ present, and thus solutions to \eqref{eq:Wind} will give eigenvalues centered at $-1$ with radius $\sigma\sqrt{SC}$, giving the same stability criteria as the single stage case \eqref{eq:1stagestabilitycriteria}. 
For an expanded derivation, see Supplementary Information. 

\begin{figure}[!htb]
\includegraphics[width=1\textwidth]{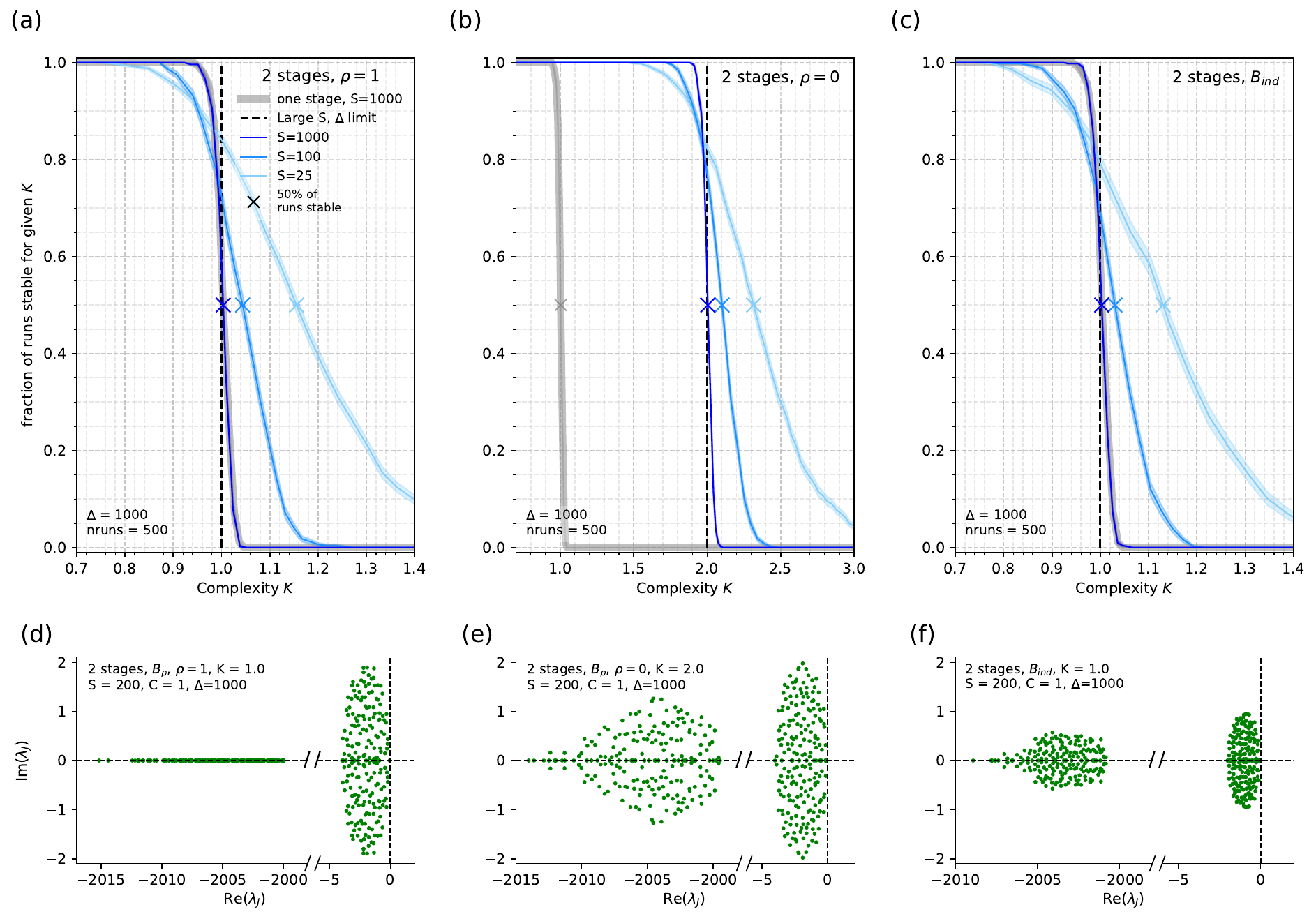}
\caption{\label{fig:Scurves}
For each model, we compute the fraction of stable realizations of the interaction matrix $\Bb$ given a specified complexity $K$ in (a, b, c). A value of $\Delta=1000$ is use to determine elements of $\Rr$ to focus on the effect of changing the complexity of the interaction matrix $\Bb$. At low $S$, there is a significant probability that a system remains linearly stable past the expected stability threshold. At large $S$, the $K$ value at which $50\%$ of realizations of matrix $\Bb$ result in stable systems approaches the asymptotic value, and the transition to instability is sharp. (a) In two-stage systems with identically interacting stages this transition occurs at $K=1.0$ and follows the classical single-stage case. (b) When interactions are independent across stages ($\rho=0$), at large $S$ the transition to instability occurs at $K=2$.  (c) When both inter-specific and cross-stage, intra-specific interactions are drawn randomly ($\Bind$), the stability transition again collapses to $K=1$ and matches the single-stage case. Standard error is computed using a bootstrapping method, and is shown by the shaded width of each curve. In (d, e, f) the eigenvalue spectra for the three models are each composed of two clusters, one near $2\Delta$ and the other determined by the form of $\Bb$.}
\end{figure}

\begin{figure}[!htb]
\includegraphics[width=1\textwidth]{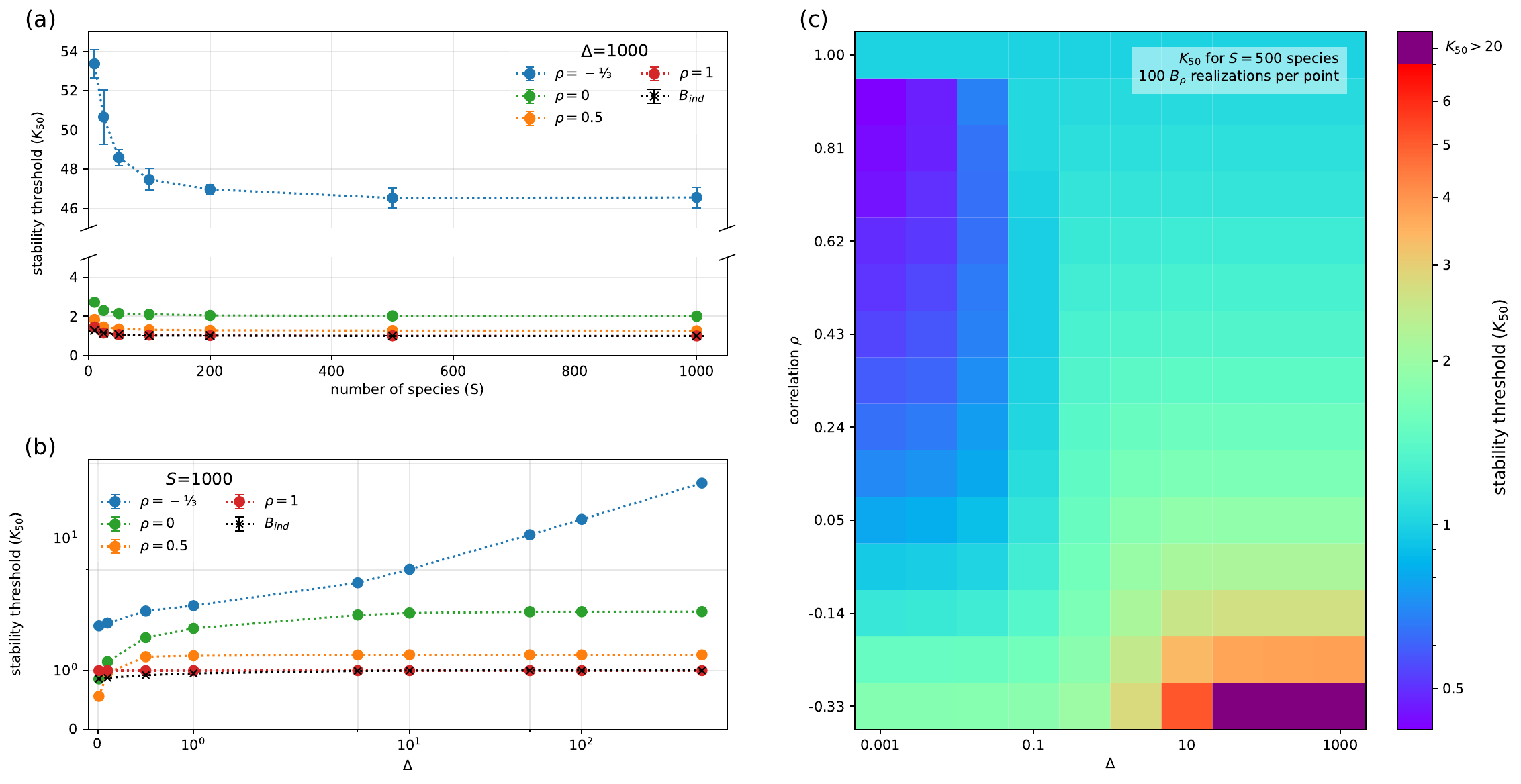}
\caption{\label{fig:deltafig} 
We numerically compute $K_{50}$, the threshold at which 50\% of random realizations of the interaction matrix remain stable across values of $S$, $\rho$, and $\Delta$. (a) At large $\Delta=1000$, dependence on system size $S$ is shown for different values of correlation $\rho$ and for the case of $\Bb_{ind}$. (b) Similarly for a large system with $S=1000$ species, dependence on $\Delta$ is shown for different values of correlation $\rho$ and for the case of $\Bb_{ind}$. At large $S$ and $\Delta$, the threshold for stability approaches the analytical prediction for each case. In (b), for decreasing correlation $\rho$, larger $\Delta$ is required to use the large $\Delta$ limit, but even at finite $\Delta$ values we see the increasing stability limit of complexity with decreased $\rho$. Standard error is computed in (a) and (b) using a bootstrapping method, but is smaller than the marker size for most points. Dashed lines are used to connect points in order to illustrate general trends. (c) Stability threshold $K_{50}$ is computed across a sweep of  $\rho$, $\Delta$ for a system of size $S=500$. While there is no $\Delta$ dependence for $\rho=1$, at large but fractional values of $\rho$ and small $\Delta$, stability is decreased as demographic rates are too small to counteract inter-specific interactions. For values of $\Delta>0.5$, decreasing $\rho$ is increasingly stabilizing. At large $\Delta$ and the minimum value of $\rho=-\frac{1}{3}$, the threshold for asymptotic stability diverges.}
\end{figure}

We confirm this result in our numerical analysis, where the eigenvalues again partition into two sets: those distributed in a circle centered at $-1$ with approximate radius $K$, and those located near $-2\Delta$, shown in Figure \ref{fig:Scurves}. We see similar behavior to the cases of $B_{\rho}$ at small $S$ or finite $\Delta$ in Figure \ref{fig:deltafig}, where the threshold of stability increases with smaller $S$ or larger $\Delta$. 

When intra-specific interactions are also drawn randomly from $N(0, \sigma^2)$, the stability threshold collapses back to the classical limit.
Intuitively, if these $2S$ stages were independent species, the larger system size would impose a penalty for higher complexity, with an increased spectral radius. However, the conversion between stages via $\Rr$ links  dynamics of different stages, effectively coarse-graining the system back into $S$ entities. 
At large $\Delta$, perturbations to the fraction of each stage within a population relax quickly relative to perturbations to the total population size.
This fast demographic mixing offsets the destabilizing effects of the larger system, returning the stability threshold exactly to that of a classical $S$-species community in the high $\Delta$ limit.

 This illustrates that to achieve the expanded stability observed in the $\rho<1$ models, two conditions must be met simultaneously: there must be functional variation in inter-specific interactions across life stages, coupled with strong, negative regulatory interactions between stages of the same species.

\begin{table}[h!]
    \caption{Results for different Interaction rules}
    \label{tab:results}
\begin{center}
\begin{tabular}{ |>{\centering\arraybackslash}m{2.5cm} | >{\centering\arraybackslash}m{3cm}|m{9cm}|}
\hline
    \textbf{Visual} &\textbf{Model}& \multicolumn{1}{c|}{\textbf{Stability Threshold for large $S, \Delta$}} \\ 
 \hline
 \adjustbox{margin=2mm}{\includegraphics[width=2.2cm]{figs/icons/onestage_black.png}}
 & Single stage ($\Aa$) & Eigenvalues are distributed in the complex plane within a circle centered at $-1$ with radius $K$. Thus, the system is stable if $\boxed{K=\sigma\sqrt{SC} <1}$ \\  
 \hline
 \adjustbox{margin=2mm}{\includegraphics[width=2.2cm]{figs/icons/rho_1_black.png}}
 & Identically interacting stages ($\Bb_1$) &  $S$ eigenvalues are distributed in a circle centered at $-2$ with radius $2K$, and $S$ eigenvalues are found at values of $-\phi_i-\gamma_i$ for $i=1:S$. Stability criteria is then identical to the single-stage case with $a_{ij}=b_{ij}$, and   $\boxed{K=\sigma\sqrt{SC} <1}$ \\  
 \hline
 \adjustbox{margin=2mm}{\includegraphics[width=2.2cm]{figs/icons/rho_08_black.png}}
 & Stage-dependent interactions ($\Bb_{\rho}$) &$S$ eigenvalues are distributd in a circle centered at $-2$ with radius $K\sqrt{1+3\rho}$, and $S$ eigenvalues are distributed near $-2\Delta$. The system is stable for $\boxed{K<\frac{2}{\sqrt{1+3\rho}}}$  \\  
 \hline
 \adjustbox{margin=2mm}{\includegraphics[width=2.2cm]{figs/icons/ind_black.png}}
 & Random intra-specific interactions ($\Bind$) & $S$ eigenvalues are disributed in a circle centered at $-1$ with radius $K$, and $S$ eigenvalues are distributed near $-2\Delta$. We return to the single-stage stability criterea of $\boxed{K=\sigma\sqrt{SC} <1}$
 \\  
 \hline
\end{tabular}
\end{center}
\end{table}

\pagebreak
\section*{Discussion}

Efforts to resolve the complexity-stability paradox in theoretical ecology have historically relied on modeling species as homogeneous populations. However, population demography fundamentally alters ecosystem processes \citep{rudolf2013}, and simple stage-structured community-assembly models can accurately reproduce population dynamics in large, complex communities \citep{rossberg2011}. When included, stage structure can mitigate the destabilizing effects of complexity \citep{mazzarisi2024}, and dynamically responsive growth and reproduction can stabilize large systems in modeled food webs \citep{Nonaka2021, deRoos2021}. Furthermore, structural biases toward certain interaction types across life stages, such as predator-prey or competitive dynamics, can shift the stability limit of an entire community \citep{gimenez2025}. 

We have incorporated explicit stage structure into a generalized Lotka-Volterra framework. In this new model, functional diversity across life stages, when coupled with continued intra-specific competition, provides a structural pathway for highly diverse communities to remain locally stable.

\subsection*{Stage structure and community stability}

Classical, single-stage models have long demonstrated that increasing the number of interacting species or their interaction variance drives a system toward instability at the strict threshold of $K<1$ \citep{may1972, Allesina_2012}. As we have shown in our $\rho=1$ case, simply partitioning a population into multiple life stages leaves this established stability criterion unchanged. It is only when demographic stages exhibit some functional diversity via ecologically heterogeneous interactions, resulting in decreased effective inter-specific interactions, that life stages increase the stability threshold. 

While our results are derived through a distinct mathematical framework, our findings resonate with other recent conclusions that population structure plays a key role in community stability \citep{gimenez2025, deRoos2021}. We demonstrate that when species are modeled with multiple life stages exhibiting distinct inter-specific interactions, large communities can remain stable up to a much higher threshold of complexity. As functional diversity is reduced, destabilizing feedback loops that can otherwise collapse large random networks are disrupted. One life stage may be strongly affected by the growth or decline of another species while other stages dampen this effect, and intra-specific functional differentiation acts as a stabilizing force for the entire community.

\subsection*{Ecological implications of functional diversity}

The spectral contraction of the Jacobian that occurs with reduced correlation $\rho$ demonstrates how the functional diversity between life stages, manifesting ecologically as ontogenetic differentiation in interaction strengths,  acts as an ecological buffer. Decreasing $\rho$ into low or even negative values represents ontogenetic niche shifts where the nature of an interaction may invert across a life cycle, and decreases the strength of the effective interactions between species. If a species experiences a highly detrimental interaction, such as heavy predation or fierce resource competition in one stage, low $\rho$ implies that this vulnerability is not carried into its other stages. For example, a juvenile may be preyed upon by a specific species (a negative interaction), but upon reaching maturity, it may outgrow this vulnerability and predate upon that same species (a positive interaction). This decoupling diffuses feedback loops that traditionally drive large, unstructured communities to instability. 

This aligns with empirical observations, where life cycles may involve abrupt and significant changes to morphology, physiology, and behavior \citep{wilbur1980}. These changes may manifest as differences in prey selection \citep{bystrom2005}, altered vulnerability to predation and parasitism \citep{briggs2000}, or complete habitat shifts \citep{schreiber2008}. Ontogenetic differentiation reduces the total overlap in resource use and predation risk for the species as a whole. Even when overall interaction variance is high, the functional decoupling in our framework allows stage populations to fluctuate somewhat independently, maintaining the species' overall resilience to community-driven perturbations.

\subsection*{Separation of timescales between interactions and demographic rates}
Our analytical results are found using the limit of large $\Delta$ to isolate the effect of the inter-specific interactions. Ecologically, this limit implies a separation of timescales: the internal life cycle  of a species operates much faster than the scale on which it interacts with other species. Thus within a species the ratio of life stages reach their relative equilibrium much faster than the aggregate populations return to (or diverges from) the community-level equilibrium. A large $\Delta$ limit allows us to effectively view each species as a single population again, with modified inter- and intra-specific interaction terms.

While our numerical results show that finite demographic rates (finite $\Delta$) reduce the asymptotic stability threshold, the stabilizing mechanism of decreased $\rho$ remains robust across $\Delta$ values (Figure \ref{fig:deltafig}(c)). When growth and fecundity rates operate on timescales comparable to inter-specific competition (of order $1$), the stability limit is still significantly increased by the functional diversity across stages. Thus, even in realistic communities with finite demographic cycles, the stabilizing effects of life stages shown here continue.

\subsection*{The necessity of intra-specific competition}

Uncorrelated inter-specific interactions across stages do not alone make a system more robust to perturbation. As demonstrated in our model with $\Bb_{ind}$, if different life stages interact with one another as distinct, uncorrelated species, the stability threshold collapses back to the classical limit of $K<1$. To achieve the expanded stability we observe in our framework, strong intra-specific competition, both within and between stages, acts as a dampening force and reduces the whole-species effective interactions.

These necessary cross-stage competitive effects can manifest as density-dependent modifications to the growth and fecundity of the species via the demographic matrix $\Rr$. In the two-stage case, as the density of juvenile stages increases, this competition can be understood as a reduction in the fecundity of the mature stage. Similarly, as the density of mature adults increases, the growth rate of juveniles into adults effectively decreases (see Supplementary Information for more details). 

This reinforces the conclusion reached by \citet{deRoos2021}: dynamically changing life cycle rates increase stability by dampening changes to the proportional distribution of a species' life stages. In nature, this is facilitated through various cross-stage mechanisms. In populations of the zooplankton \textit{Daphnia} in lakes, reproduction of viable offspring is greatly reduced when food is scarce due to intra-specific competition \citep{mccauley1999}. In many amphibian species, high larval density in aquatic environments restricts future adult size and initiates a delayed life-history effect that significantly reduces the terrestrial adult's reproduction rate \citep{scott1994}. Beyond direct competition, density-dependent regulation can also be mediated by third-species effects, classic to the Janzen-Connell effect \citep{janzen1970, connell1971}. By attracting host-specific natural enemies, a high density of adult trees actively suppresses the maturation of the more vulnerable seedlings of the same species nearby \citep{mangan2010}. 

\subsection*{Limitations and future directions}

While this modified generalized Lotka-Volterra framework gives critical insights into the role of population stage structure in community stability, it relies on mathematical simplifications that warrant further analytical exploration. We strictly examined fully random interaction networks with interactions drawn from symmetrical distributions; however, real communities may often exhibit biases toward different interaction types, including predator-prey interactions, inter-species competition, or mutualism \citep{Allesina_2012}. 

Additionally, for a guaranteed feasible equilibrium, our modified Lotka-Volterra framework is under-constrained and relies on a choice or values within the demographic matrix $\Rr$. While we mitigate this using $\Delta$, more complex methods for assigning demographic rates may be explored, where juvenile and adult rates may be coupled.

To isolate the stabilizing mechanisms of stage structure, we currently rely on a binary contrast: cross-stage regulation is either fixed as strongly negative or drawn randomly. In natural systems, intra-specific competition can exist in a broad continuum, and be weakly negative, asymetric, or context-dependent. A critical future direction is the analysis of different forms of intra-speicific quadratic interactions, through a continuous parameter controlling the magnitude of these negative interactions or a more complex formulation. Mapping how the stability threshold shifts from weak to strong regulation can provide a clearer connection between the mathematical bounds we utilized and the nuanced reality of biological communities.

Beyond theoretical extensions, the most pressing future direction is interfacing these mathematical predictions with empirical data. Accurately characterizing interactions in large, multi-stage systems in wild communities remains observationally challenging, but experimental microcosms offer a promising path forward. Recent work in characterizing the complexity of assembled microbial communities has successfully demonstrated May’s instability transition \textit{in vitro} \citep{ratzke2020, hu2022}. Although microbes do not exhibit classical juvenile-adult ontogeny, applying our block-structured matrix framework to similar experimental systems, where these stages instead represent different phenotypic states or metabolic phases, could allow for the direct empirical testing of how intra-specific functional differentiation alters the stability-complexity threshold.

Ultimately, the framework presented here gives fundamental insight into how stage structure and population heterogeneity modify the stability criterion of a community. By demonstrating that functionally diverse life stages buffer against the destabilizing effect of increasing system size and complexity, this work bridges demographic life-history theory with random matrix models, and provides a theoretical foundation for understanding how ontogeny expands the complexity boundaries within which modeled communities can remain stable.

\pagebreak

\section*{Acknowledgments}

We thank our colleagues in the Plant Biology and Physics departments at the University of Illinois, Urbana-Champaign for continued support and feedback on this project. 

G.G. thanks Douglas Domingues for helpful and insightful discussions during this process. G.G. thanks Damla Cinoglu for guidance during writing and revising of this manuscript. 

J.P.OD. acknowledges National Science Foundation, Division of Environmental Biology (grant 2401124).

\pagebreak
\bibliographystyle{ecology_letters2}
\bibliography{bib/gracerefs}{}

\end{document}



\begin{flushleft}
{\Large
\textbf{}
}
\\
Grace Garmire $^{1}$
James P. O'Dwyer  $^{2}$
\\
\bf{1} Department of Physics, University of Illinois, Urbana IL USA \\
\bf{2} Department of Plant Biology, University of Illinois, Urbana IL USA \\

\section*{Supplementary Information for: Heterogeneity among and competition between life stages together extend the stability boundary of complex \ed{ecological communities}}

\end{flushleft}

\subsection*{\includegraphics[height=4em]{figs/icons/onestage_black.png} One stage analytical solution}

Following the work of \citet{may1972}, we want to see the behavior of a non-linear system close to an equilibrium. For a system of $S$ interacting species, the growth rate of species $i$ is a function of all other species: \begin{align}
    \frac{dx_i}{dt}=f_i(x_1, x_2, \dots , x_S)
\end{align} 

In the classic gLV case we have \begin{align}
    \frac{dx_i}{dt}=r_ix_i+x_i\sum_{j=1}^s A_{ij}x_j
\end{align}\begin{align}
    \Aa = \begin{bmatrix}
        -1 & a_{12} & a_{13} & a_{14} & \cdots \\
        a_{21} & -1 & a_{23} & a_{24} & \cdots \\
        a_{31} & a_{32} & -1 & a_{34} & \cdots \\
        a_{41}& a_{42}& a_{43} & -1 & \cdots  \\ \vdots & \vdots & \vdots & \vdots & \ddots
    \end{bmatrix}
\end{align} where $\vec{r}$ is a vector of growth rates and matrix $A_{ij}$ is a random matrix with diagonals of $-1$, and off-diagonal entries are nonzero with probability $C$ and then drawn from the normal distribution $N(0, \sigma^2)$. 

At a feasible equilibrium $\vec{x}^*$, for every species $\frac{dx_i}{dt}=0$. To understand if the system is stable, we look at perturbations away from that equilibrium: $x_i(t)=x_i^*+y_i(t)$. Taking only first-order terms, this simplifies to a linear equation: \begin{align}
    \frac{d\vec{y}}{dt}= \Aa \vec{y}
\end{align} This has the same effect as fixing our stable equilibrium at $\vec{x}^*=\vec{1}$.

The general solution to this system is a linear combination of terms from the eigenvalues of the equilibrium Jacobian, \begin{align}
    y_i(t)=\sum_{j=1}^Sc_je^{\lambda_jt} v_j
\end{align}

We investigate the Jacobian at this equilibrium to determine the local stability of the system, \ed{$J_{ij}^*=\left.\frac{d\dot{x}_i}{dx_j}\right|_{\vec{x}^*=\vec{1}}$} The Jacobian at equilibrium is then \begin{align}
    \Js = \Aa
\end{align} and stability is dependent on the eigenvalues of the interaction matrix. As a random matrix, $\Aa$ has a well-understood form. Using Wigner's Circular Law from random matrix theory, we know the eigenvalues of $\Aa$ will form a disk in the complex plane with radius $K=\sigma\sqrt{SC}$, centered at $-1$ on the real axis \citep{may1972, tao2010}.
Thus, the leading eigenvalue will have negative real part for $\sigma\sqrt{SC}<1$, giving our stability criteria.

\subsection*{Analytical solutions for two life stages}

In our stage structured model, our interaction matrix \ed{for a two-stage system} is \ed{\begin{align}
    \Bb=\begin{bmatrix} 
-1 & b_{11}^{12} & b_{12}^{11} & b_{12}^{12} & \cdots \\
b_{11}^{21} & -1 & b_{12}^{21} & b_{12}^{22} & \cdots \\
b_{21}^{11} & b_{21}^{12} & -1 & b_{22}^{12} & \cdots \\
b_{21}^{21} & b_{21}^{22} & b_{22}^{21} & -1 & \cdots \\
\vdots & \vdots & \vdots & \vdots &\ddots
\end{bmatrix}
\end{align}}
and \ed{the growth rate vector $\vec{r}$ is replaced by a demographic matrix,} \begin{align}\label{ap:Rdef}
    \Rr=\begin{bmatrix}
\mu_1^1 & \phi_1 & 0 & 0& \cdots \\ \gamma_1 & \mu_1^2 & 0 & 0& \cdots \\
0&0 & \mu_2^1 & \phi_2 & \cdots\\ 0&0 & \gamma_2 & \mu_2^2 & \cdots\\ \vdots&\vdots &\vdots &\vdots & \ddots
\end{bmatrix}
\end{align}
Our system is now governed by:
\begin{equation}\label{ap:2stagesystem}\frac{dx^{\alpha}_{i}}{dt}=\sum_{j,\beta}R_{ij}^{\alpha\beta}x_{j}^\beta+x_{i}^\alpha\sum_{j,\beta} b_{ij}^{\alpha\beta}x_{j}^\beta
\end{equation}
\ed{Where the summation is taken over every stage $(\beta=1:L)$ of every species $(j=1:S)$ to capture all $S\cdot L$ interactions that affect stage $\alpha$ of species $i$.}

At equilibrium, $\vec{x} = \vec{x}^*$ and $\frac{d\vec{x}}{dt}|_{\vec{x}^*} = 0$. We can again rescale population values so each population density is measured relative to its equilibrium value, resulting in an equilibrium at $\vec{x}^*=\vec{1}$. Assuming a feasible equilibrium exists does not affect the \ed{asymptotic} stability result of the system \citep{gibbs2018}. \ed{Putting this into \eqref{ap:2stagesystem} }results in the equilibrium condition of \begin{align}\label{ap:2stageeq}
    \sum_{j,\beta}b_{ij}^{\alpha\beta}=-\sum_{j,\beta}R_{ij}^{\alpha\beta}
\end{align}
\ed{Elements of the Jacobian of the two-stage system are found from equation \eqref{ap:2stagesystem} as \begin{align}
    J_{ij}^{\alpha\beta}&=\frac{d}{dx_j^\beta} \left(\frac{dx_i^\alpha}{dt}\right) \\ J_{ij}^{\alpha\beta} &= R^{\alpha\beta}_{ij}+x^{\alpha}_i b^{\alpha\beta}_{ij}+\delta_{ij}\delta_{\alpha\beta}\sum_{k,\xi}b_{ik}^{\alpha\xi}
\end{align}
At the equilibrium of $\vec{x}^*=\vec{1}$, \begin{equation}
    {J^*}_{ij}^{\alpha\beta} = R_{ij}^{\alpha\beta}+b_{ij}^{\alpha\beta}+\delta_{ij}\delta_{\alpha\beta}
\sum_{k,\xi}b_{ik}^{\alpha\xi}\end{equation} Using \eqref{ap:2stageeq}, \begin{align}\label{ap:2stagejac}
    {J^*}_{ij}^{\alpha\beta} &= R_{ij}^{\alpha\beta}+B_{ij}^{\alpha\beta}-\delta_{ij}\delta_{\alpha\beta}\sum_{k,\xi}R_{ik}^{\alpha\xi}
\end{align}
From \eqref{ap:Rdef}, we can see \begin{align}
    \sum_{k,\xi}R_{ik}^{\alpha\xi}=\begin{cases}
        \mu_i^1+\phi_i\ & \alpha=1 \\\gamma_i+\mu_i^2\ & \alpha=2
    \end{cases}
\end{align}
The first and third terms of \eqref{ap:2stagejac} now only involve terms in $\Rr$, and can be combined to the simpler form $\Rr'=\Rr - \diag{\sum_{k,\xi}R_{ik}^{\alpha\xi}}$. This is a direct result of the feasibility constraining values in $\Rr$, and thus cancelling out the rowsums of $\Bb$. }

The Jacobian at equilibrium  \ed{then
\begin{align}
    \Js &= \Rr'+\Bb \end{align}\begin{align}
    \Js &= \underbrace{\begin{bmatrix}
-\phi_1 & \phi_1 &0&0&\cdots\\ \gamma_1 & -\gamma_1 &0&0&\cdots\\ 0&0 & -\phi_2 & \phi_2 &\cdots\\ 0&0 & \gamma_2 & -\gamma_2 &\cdots\\  \vdots&\vdots &\vdots &\vdots & \ddots
\end{bmatrix}}_{\Rr'} + \underbrace{\begin{bmatrix}
    -1 & b_{11}^{12} & b_{12}^{11} & b_{12}^{12} & \cdots \\
b_{11}^{21} & -1 & b_{12}^{21} & b_{12}^{22} & \cdots \\
b_{21}^{11} & b_{21}^{12} & -1 & b_{22}^{12} & \cdots \\
b_{21}^{21} & b_{21}^{22} & b_{22}^{21} & -1 & \cdots \\
\vdots & \vdots & \vdots & \vdots &\ddots
\end{bmatrix}}_{\Bb}
\end{align}}
We again need to find the eigenvalues of this Jacobian at equilibrium to determine the local stability. The characteristic equation of $\Jj$ comes from $\det(\Js-\lambda \Id)=0$: \ed{\begin{align}\label{ap:det2jac}
    0=\det\left(\Js-\lambda\Id\right) = 
    \begin{vmatrix}
        -1-\phi_1-\lambda & b_{11}^{12}+\phi_1 & b_{12}^{11} & b_{12}^{12} & \cdots \\
b_{11}^{21}+\gamma_1 & -1-\gamma_1-\lambda & b_{12}^{21} & b_{12}^{22} & \cdots \\
b_{21}^{11} & b_{21}^{12} & -1-\phi_2-\lambda & b_{22}^{12}+\phi_2 & \cdots \\
b_{21}^{21} & b_{21}^{22} & b_{22}^{21}+\gamma_2 & -1-\gamma_2-\lambda & \cdots \\
\vdots & \vdots & \vdots & \vdots &\ddots
    \end{vmatrix} 
\end{align}}The determinant of a matrix is invariant under the addition or subtraction of any row (or column) to another row (column). The determinant is multiplied by a factor of $-1$ by the swapping of any two rows (columns) \citep{Abadir2005}. \ed{We perform the following series of these elementary row operations on \eqref{ap:det2jac}:\begin{enumerate}
    \item For every $m$ odd, add column $m+1$ to column $m$ (leaving column $m+1$ unchanged).
    \item For every $n$ odd, subtract row $n$ from row $n+1$ (leaving row $n$ unchanged).
    \item Reorder rows as $r_1, r_3, r_5, \dots, r_2, r_4, r_6, \dots $
    \item Reorder columns as $c_1, c_3, c_5, \dots, c_2, c_4, c_6, \dots $
\end{enumerate}}
The determinant remains unchanged for operations $1$ and $2$, and operations $3$ and $4$ result in an even number of $S(S-1)$ row changes, \ed{also leaving the determinant unchanged, so our equivalent expression is now}:\ed{\begin{align}\label{ap:det2jac_new}
    &\det\left(\Js-\lambda\Id\right) =\nonumber\\&\quad \begin{vmatrix}
       -1+b_{11}^{12}-\lambda & b_{12}^{11}+b_{12}^{12} & \cdots & b_{11}^{12}+\phi_1 & b_{12}^{12} & \cdots \\ b_{21}^{11} + b_{21}^{12} & -1+b_{22}^{12}-\lambda & \cdots & b_{21}^{12} & b_{22}^{12}+\phi_2 & \cdots \\ 
       \vdots & \vdots & \ddots & \vdots & \vdots &\ddots \\
       b_{11}^{21}-b_{11}^{12} & b_{12}^{21}+b_{22}^{12}-b_{12}^{11}-b_{12}^{12} & \cdots & -1-b_{11}^{12}-\gamma_1-\phi_1-\lambda & b_{12}^{22}-b_{12}^{12} & \cdots \\ 
       b_{21}^{21}+b_{21}^{22}-b_{21}^{11}-b_{21}^{12} & b_{22}^{21}-b_{22}^{12} & \cdots & b_{21}^{22}-b_{21}^{12} & -1-b_{22}^{12}-\gamma_1-\phi_1-\lambda  & \cdots\\
\vdots & \vdots & \vdots & \vdots&  \vdots &\ddots
    \end{vmatrix} 
\end{align}
We start from here to determine the eigenvalues of each form of our interaction matrix below.
}

\subsubsection*{\includegraphics[height=4em]{figs/icons/rho_1_black.png} $\rho=1$: Identically interacting stages}
\ed{When stages interact identically, $\rho=1$ so $b_{ij}^{\alpha\beta}=b_{ij}$ and $b_{ii}^{\alpha\beta}=-1$ for all $\alpha,\beta$. Now \eqref{ap:det2jac_new} becomes }\begin{align}\label{ap:det2jac_p1}
    \det\left(\Js_{1}-\lambda\Id\right) = \begin{vmatrix}
       -2-\lambda & 2b_{12} & \cdots & -1+\phi_1 & b_{12} & \cdots \\ 2b_{21} & -2-\lambda & \cdots & b_{21} & -1+\phi_2 & \cdots \\
       \vdots & \vdots & \ddots & \vdots & \vdots &\ddots \\
       0 & 0 & \cdots & -\gamma_1-\phi_1-\lambda & 0 & \cdots \\ 0 & 0 & \cdots & 0 & -\gamma_2-\phi_2-\lambda & \cdots \\
\vdots & \vdots & \vdots & \vdots & \vdots &\ddots
    \end{vmatrix} 
\end{align}
This matrix is neatly divided into 4 \ed{sub-matrices. For such a partitioned matrix, we have \citep{Abadir2005}}: \begin{align}
    \begin{vmatrix}
        \mathbf{W} & \mathbf{X} \\ \mathbf{0} & \mathbf{Z}
    \end{vmatrix} = \det(\mathbf{W})\det(\mathbf{Z})
\end{align} The eigenvalues are then distributed in two distinct \ed{sets from the solutions  $\det(\mathbf{X})=0$ and  $\det(\mathbf{Z})=0$}: \begin{align}
    0&=\det(\mathbf{X}) = \begin{vmatrix}
       -2-\lambda_B & 2b_{12}& 2b_{13} & \cdots \\ 2b_{21} & -2-\lambda_B & 2b_{23}& \cdots \\2b_{31} & 2b_{32} & -2-\lambda_B & \cdots \\
       \vdots & \vdots & \vdots & \ddots 
    \end{vmatrix} \end{align}\begin{align}
    0&=\det(\mathbf{Z}) = \begin{vmatrix}
        -\gamma_1-\phi_1-\lambda_{R'} & 0 & 0 & \cdots \\ 0& -\gamma_2-\phi_2-\lambda_{R'} & 0 &  \cdots \\ 0& 0& -\gamma_3-\phi_3-\lambda_{R'} & \cdots  \\ \vdots& \vdots &\vdots& \ddots
    \end{vmatrix}
\end{align} 
The first set of eigenvalues, $\{\lambda_B\}$, are distributed in a circle centered at $-2$ with a radius of $2\sigma\sqrt{SC}=2K$ that from Wigner's Circular Law. The second set is entirely composed of elements in $\Rr$: $\{\lambda_{R'}\}=\{-\phi_i-\gamma_i\}$. Since biological constraints on the model require $\phi_i, \gamma_i\geq 0$, any positive leading eigenvalue of $\Js$ will come from the set of $\{\lambda_B\}$ and thus the stability criterion is identical to the one-stage case: $K<\sigma\sqrt{SC}$. 

\subsubsection*{\includegraphics[height=4em]{figs/icons/rho_08_black.png} $\rho \neq1:$ varied interactions}

When interactions are not identical across stages, interactions between two species vary by stage such that $\cov{b_{ij}^{\alpha\beta}, b_{ij}^{\alpha'\beta'}}=\rho\sigma^2$, \ed{but again $b_{ii}^{\alpha\beta}=-1$ for all $\alpha,\beta$. Using \eqref{ap:det2jac_new}, now 
\begin{align} \det\left(\Js_{\rho}-\lambda\Id\right) =
    \begin{vmatrix}
       -2-\lambda & b_{12}^{11}+b_{12}^{12} & \cdots & -1+\phi_1 & b_{12}^{12} & \cdots \\ b_{21}^{11} + b_{21}^{12} & -2-\lambda & \cdots & b_{21}^{12} & -1+\phi_2 & \cdots \\ 
       \vdots & \vdots & \ddots & \vdots & \vdots &\ddots \\
       0 & b_{12}^{21}+b_{22}^{12}-b_{12}^{11}-b_{12}^{12} & \cdots & -\gamma_1-\phi_1-\lambda & b_{12}^{22}-b_{12}^{12} & \cdots \\ 
       b_{21}^{21}+b_{21}^{22}-b_{21}^{11}-b_{21}^{12} & 0 & \cdots & b_{21}^{22}-b_{21}^{12} & -\gamma_1-\phi_1-\lambda  & \cdots\\
\vdots & \vdots & \vdots & \vdots&  \vdots &\ddots
    \end{vmatrix} 
\end{align}
At the limit of large $\Delta$, rates $\phi, \gamma=\Delta$ and \begin{align}\label{ap:largedeltaJ-lambdaI}
    \lim_{\Delta\to\infty}{\det\left(\Js_{\rho}-\lambda\Id\right)} = 
    \begin{vmatrix}
       -2-\lambda & b_{12}^{11}+b_{12}^{12} & \cdots & -1+\Delta & b_{12}^{12} & \cdots \\ b_{21}^{11} + b_{21}^{12} & -2-\lambda & \cdots & b_{21}^{12} & -1+\Delta & \cdots \\ 
       \vdots & \vdots & \ddots & \vdots & \vdots &\ddots \\
       0 & b_{12}^{21}+b_{22}^{12}-b_{12}^{11}-b_{12}^{12} & \cdots & -2\Delta-\lambda & b_{12}^{22}-b_{12}^{12} & \cdots \\ 
       b_{21}^{21}+b_{21}^{22}-b_{21}^{11}-b_{21}^{12} & 0 & \cdots & b_{21}^{22}-b_{21}^{12} & -2\Delta-\lambda  & \cdots\\
\vdots & \vdots & \vdots & \vdots&  \vdots &\ddots
    \end{vmatrix} 
\end{align}
This is again a partitioned matrix. If $\mathbf{Z}$ is nonsingular we have 
\begin{align}\label{ap:Znonsing}
    \begin{vmatrix}
        \mathbf{W} & \mathbf{X} \\ \mathbf{Y} & \mathbf{Z}
    \end{vmatrix} = \det(\mathbf{Z})\det(\mathbf{W-XZ^{-1}Y})
\end{align} and if $\mathbf{W}$ is nonsingular we have \begin{align}\label{ap:Wnonsing}
    \begin{vmatrix}
        \mathbf{W} & \mathbf{X} \\ \mathbf{Y} & \mathbf{Z}
    \end{vmatrix} = \det(\mathbf{W})\det(\mathbf{Z-YW^{-1}X})
\end{align}}
\ed{
We can use \eqref{ap:Znonsing} while $\det(\mathbf{Z}(\lambda_J))\neq 0$, ensuring $\mathbf{Z}$ is nonsingular. Thus half of our eigenvalues of $\Jj$ come from solving $\det(\mathbf{W-XZ^{-1}Y})=0$.}
\ed{
We can rewrite the sub-matrix $\mathbf{Z}$ as \begin{align}
    \mathbf{Z}\ = (-2\Delta-\lambda)\Id + \mathbf{N} \qquad \text{with }   \mathbf{N} =\begin{bmatrix}
        0& b_{12}^{22}-b_{12}^{12} & b_{13}^{22}-b_{13}^{12}& \cdots \\ b_{21}^{22}-b_{21}^{12} & 0& b_{23}^{22}-b_{23}^{12}& \cdots \\
         b_{31}^{22}-b_{31}^{12} & b_{32}^{22}-b_{32}^{12}& 0& \cdots \\ \vdots & \vdots & \vdots & \ddots 
    \end{bmatrix}
\end{align}
Then \begin{align}
    \mathbf{Z}^{-1} = \left((-2\Delta-\lambda) \Id + \mathbf{N} \right)^{-1} = \frac{-1}{2\Delta+\lambda}\left(\Id-\frac{1}{2\Delta+\lambda}\mathbf{N}\right)
\end{align}Now $\mathbf{Z}^{-1}$ looks like a Neumann series, where for a linear bounded operator $T$, if $(I-T)$ is invertible, its inverse is \begin{align}
    (\Id-\mathbf{T})^{-1} = \sum_{k=0}^\infty \mathbf{T}^k
\end{align} \citep{kreyszig1989}. For $\mathbf{Z}$ we have \begin{align}
    \mathbf{T}=\frac{1}{2\Delta+\lambda}\mathbf{N}
\end{align} which is linear and bounded. So \begin{align}
    \mathbf{Z}^{-1} = \frac{-1}{2\Delta-\lambda}\sum_{k=0}^\infty \mathbf{N}^k = \frac{-1}{2\Delta+\lambda}\Id + \frac{1}{(2\Delta+\lambda)^2}\mathbf{N} - \frac{1}{(2\Delta+\lambda)^3}\mathbf{N}^2+\dots
\end{align}
In the limit of large $\Delta$, we keep only the first term, and drop terms of order $\mathcal{O}(\Delta^{-2})$ and smaller: 
\begin{align}
    \mathbf{Z}^{-1}=\frac{-1}{2\Delta+\lambda} \Id + \mathcal{O}(\Delta^{-2})
\end{align}
and after dropping terms of order $\mathcal{O}(\Delta^{-2})$ and smaller at large $\Delta$, our expression inside the determinant is 
\begin{align}
    \lim_{\Delta\to\infty}(\mathbf{W-XZ^{-1}Y})=&\begin{bmatrix}
        -2-\lambda & b_{12}^{11}+b_{12}^{12} & \cdots \\
        b_{21}^{12}+b_{21}^{12} & -2-\lambda & \cdots \\
        \vdots & \vdots & \ddots 
    \end{bmatrix}\nonumber \\&-\frac{-1}{2\Delta+\lambda}\begin{bmatrix}
        -1+\Delta & b_{12}^{12}& \cdots \\ b_{21}^{12} & -1+\Delta & \cdots \\ \vdots & \vdots & \ddots 
    \end{bmatrix}
    \begin{bmatrix}
        0 & b_{12}^{21}+b_{12}^{22}-b_{12}^{11}-b_{12}^{12} & \cdots \\ b_{21}^{21}+b_{21}^{22}-b_{21}^{11}-b_{21}^{12} & 0 & \cdots \\ \vdots & \vdots & \ddots 
    \end{bmatrix}
\end{align}
\begin{align}
    \lim_{\Delta\to\infty}(\mathbf{W-XZ^{-1}Y}) &= \begin{bmatrix}
        -2-\lambda & b_{12}^{11}+b_{12}^{12} & \cdots \\
        b_{21}^{12}+b_{21}^{12} & -2-\lambda & \cdots \\
        \vdots & \vdots & \ddots 
    \end{bmatrix} +\frac{-1+\Delta}{2\Delta+\lambda}\begin{bmatrix}
        0 & b_{12}^{21}+b_{12}^{22}-b_{12}^{11}-b_{12}^{12} & \cdots \\
        b_{21}^{21}+b_{21}^{22}-b_{21}^{11}-b_{12}^{12} & 0 & \cdots \\
         \vdots & \vdots & \ddots  
    \end{bmatrix}\end{align}
    From the diagonal elements, we see eigenvalues will be near $-2$ on the real axis, so we can take $\frac{-1+\Delta}{2\Delta+\lambda}\to\frac{1}{2}$, and 
    \begin{align}
    \lim_{\Delta\to\infty}(\mathbf{W-XZ^{-1}Y}) &= \begin{bmatrix}
        -2-\lambda & \frac{1}{2}\left(b_{12}^{21}+b_{12}^{22}+b_{12}^{11}+b_{12}^{12}\right) & \cdots \\
        b_{21}^{21}+b_{21}^{22}+b_{21}^{11}+b_{12}^{12} & -2-\lambda &  \cdots \\
         \vdots & \vdots & \ddots  
        \end{bmatrix}\label{ap:matrixforsmalleigs}
\end{align}
Which has off-diagonals with variance $\text{Var}\left(\frac{1}{2}\left(b_{ij}^{21}+b_{ij}^{22}+b_{ij}^{11}+b_{ij}^{12}\right)\right)=(1+3\rho)\sigma^2$. Thus, from Wigner's circle law, this set of eigenvalues from solving this characteristic equation will be distributed in a circle centered at $-2$ with radius $\sigma\sqrt{SC(1+3\rho)}$.
}

\ed{
The other half of our eigenvalues can be found using \eqref{ap:Wnonsing} while $\det(\mathbf{W})\neq 0$, by solving $\det(\mathbf{Z-YW^{-1}X})=0$. Following the procedure as above: 
\begin{align}
    \mathbf{W}^{-1}=\frac{-1}{2+\lambda} \Id + \mathcal{O}(\lambda^{-2})
\end{align}
\begin{align}
    \lim_{\Delta\to\infty}(\mathbf{Z-YW^{-1}X}) = &\begin{bmatrix}
        -2\Delta-\lambda & b_{12}^{22}-b_{12}^{12} & \cdots & \\
        b_{21}^{22}-b_{21}^{12} & -2\Delta-\lambda  & \cdots \\ \vdots & \vdots & \ddots
    \end{bmatrix}\nonumber\\
    &-\frac{-1}{2+\lambda}\begin{bmatrix}
        0 & b_{12}^{21}+b_{12}^{22}-b_{12}^{11}-b_{12}^{12} & \cdots \\ b_{21}^{21}+b_{21}^{22}-b_{21}^{11}-b_{21}^{12} & 0 & \cdots \\ \vdots & \vdots & \ddots 
    \end{bmatrix}
    \begin{bmatrix}
        -1+\Delta & b_{12}^{12}& \cdots \\ b_{21}^{12} & -1+\Delta & \cdots \\ \vdots & \vdots & \ddots 
    \end{bmatrix}
\end{align}
\begin{align}
    \lim_{\Delta\to\infty}(\mathbf{Z-YW^{-1}X}) =& \begin{bmatrix}
        -2\Delta-\lambda & b_{12}^{22}-b_{12}^{12} & \cdots & \\
        b_{21}^{22}-b_{21}^{12} & -2\Delta-\lambda  & \cdots \\ \vdots & \vdots & \ddots
    \end{bmatrix} \nonumber\\&+ \frac{-1+\Delta}{2+\lambda}\begin{bmatrix}
        0 & b_{12}^{21}+b_{12}^{22}-b_{12}^{11}-b_{12}^{12} & \cdots \\
        b_{21}^{21}+b_{21}^{22}-b_{21}^{11}-b_{21}^{12} & 0 & \cdots \\ \vdots & \vdots &  \ddots 
    \end{bmatrix}\label{ap:loweigs}
\end{align}
Where in the last line we drop all terms of order $\mathcal{O}(\Delta^{0})$ and smaller.
}

\ed{From \eqref{ap:loweigs} we can see that the eigenvalues from solving $\det(\mathbf{Z-YW^{-1}X})=0$ for large $\Delta$ will be near $-2\Delta$. Taking $\frac{-1+\Delta}{2+\lambda} \to -\frac{1}{2}$, we can further simplify \eqref{ap:loweigs}  to 
\begin{align}
    \lim_{\Delta\to\infty}(\mathbf{Z-YW^{-1}X}) =&\begin{bmatrix}
        -2\Delta-\lambda & \frac{1}{2}\left(b_{12}^{11}+b_{12}^{22}-b_{12}^{12}-b_{12}^{21}\right) & \cdots \\ 
        \frac{1}{2}\left(b_{21}^{11}+b_{21}^{22}-b_{21}^{12}-b_{21}^{21}\right) & -2\Delta-\lambda & \cdots \\ \vdots & \vdots & \ddots 
    \end{bmatrix}\label{ap:lim_largeeigs}
\end{align}
From \eqref{ap:lim_largeeigs}, eigenvalues will be distributed in a circle centered at $-2\Delta$ with radius $\sigma\sqrt{SC(1-\rho)}$. }

\ed{Even for large but finite $\Delta$, we see these two distinct sets of eigenvalues, and thus the condition of one of the diagonal blocks being non-singular is met for each set. Since the more negative eigenvalues are near $-2\Delta$ (verified numerically in \ref{M-fig:deltafig}), the leading eigenvalue always comes from \eqref{ap:matrixforsmalleigs}, and we are left with the stability criteria at large $S$ and $\Delta$ of $K<\frac{2}{\sqrt{1+3\rho}}$.
}

\subsubsection*{\includegraphics[height=4em]{figs/icons/ind_black.png} Random intra-specific interactions}

\ed{When intra-specific interactions as well as inter-specific interactions are random, the interaction matrix $\Bind$ is a fully random matrix with $-1$ diagonal entries and all other entries independently drawn from $N(0,\sigma^2)$. Now $b_{ii}^{\alpha\beta}$ is drawn randomly for  for $\alpha\neq \beta$ and all random entries are uncorrelated. Using \eqref{ap:det2jac_new} with $\phi_i, \gamma_i=\Delta$ and the same partitioned matrix procedure as above, we have 
\begin{align}
    \det(\Js_{ind}-\lambda \Id) &= \begin{vmatrix}
        -1+b_{11}^{12}-\lambda & b_{12}^{11}+b_{12}^{12} & \cdots & b_{11}^{12}+\Delta & b_{12}^{12} & \cdots \\
        b_{21}^{11}+b_{21}^{12} & -1+b_{22}^{12}-\lambda & \cdots &b_{12}^{21} & b_{22}^{21}+\Delta & \cdots \\
        \vdots & \vdots & \ddots & \vdots & \vdots & \ddots \\
        b_{11}^{21}-b_{11}^{12} & b_{12}^{21}+b_{12}^{22}-b_{12}^{11}-b_{12}^{12} & \cdots & -1-b_{11}^{12}-2\Delta-\lambda & b_{12}^{22}-b_{12}^{12} & \cdots \\
        b_{21}^{21}+b_{21}^{22}-b_{21}^{11}-b_{21}^{12} & b_{22}^{21}-b_{22}^{12} & \cdots & b_{21}^{22}-b_{21}^{12} & -1-b_{22}^{12}-2\Delta-\lambda & \cdots \\
        \vdots & \vdots & \ddots & \vdots & \vdots & \ddots \\
    \end{vmatrix}
\end{align}So
\begin{align}
    \mathbf{Z} = \begin{bmatrix}
        -1-b_{11}^{12}-2\Delta -\lambda & b_{12}^{22}-b_{12}^{12} & \cdots \\
        b_{21}^{22}-b_{21}^{12} & -1-b_{22}^{12}-2\Delta-\lambda & \cdots \\
        \vdots & \vdots & \ddots 
    \end{bmatrix} = (-1-2\Delta-\lambda)\Id + \underbrace{\begin{bmatrix}
        -b_{11} & b_{12}^{22}-b_{12}^{12} & \cdots \\
        b_{21}^{22}-b_{21}^{12} & -b_{22}^{12} & \cdots \\
        \vdots & \vdots & \ddots 
    \end{bmatrix}}_{\mathbf{N}}
\end{align}
At large $\Delta$,
\begin{align}
    \mathbf{Z}^{-1} = \frac{-1}{2\Delta+1+\lambda}\Id
\end{align}
\begin{align}
    \lim_{\Delta\to\infty}(\mathbf{W-XZ^{-1}Y})=&\begin{bmatrix}
        -1+b_{11}^{12}-\lambda & b_{12}^{11}+b_{12}^{12} & \cdots \\
        b_{21}^{12}+b_{21}^{12} & -1+b_{22}^{12}-\lambda & \cdots \\
        \vdots & \vdots & \ddots 
    \end{bmatrix}\nonumber \\&-\frac{-1}{2\Delta+1+\lambda}\begin{bmatrix}
        b_{11}^{12}+\Delta & b_{12}^{12}& \cdots \\ b_{21}^{12} & b_{22}^{12}+\Delta & \cdots \\ \vdots & \vdots & \ddots 
    \end{bmatrix}
    \begin{bmatrix}
        b_{11}^{21}-b_{11}^{12} & b_{12}^{21}+b_{12}^{22}-b_{12}^{11}-b_{12}^{12} & \cdots \\ b_{21}^{21}+b_{21}^{22}-b_{21}^{11}-b_{21}^{12} & b_{22}^{21}-b_{22}^{12} & \cdots \\ \vdots & \vdots & \ddots 
    \end{bmatrix}
\end{align}
\begin{align}
    \lim_{\Delta\to\infty}(\mathbf{W-XZ^{-1}Y})=&\begin{bmatrix}
        -1+b_{11}^{12}-\lambda & b_{12}^{11}+b_{12}^{12} & \cdots \\
        b_{21}^{12}+b_{21}^{12} & -1+b_{22}^{12}-\lambda & \cdots \\
        \vdots & \vdots & \ddots 
    \end{bmatrix}\nonumber\\
    &+\frac{\Delta}{2\Delta+1+\lambda}\begin{bmatrix}
        b_{11}^{21}-b_{11}^{12} & b_{12}^{21}+b_{12}^{22}-b_{12}^{11}-b_{12}^{12} & \cdots \\ b_{21}^{21}+b_{21}^{22}-b_{21}^{11}-b_{21}^{12} & b_{22}^{21}-b_{22}^{12} & \cdots \\ \vdots & \vdots & \ddots 
    \end{bmatrix}
\end{align}
\begin{align}
    \lim_{\Delta\to\infty}(\mathbf{W-XZ^{-1}Y})=& \begin{bmatrix}
        -1+\frac{1}{2}(b_{11}^{21}-b_{11}^{12})-\lambda & \frac{1}{2}\left(b_{12}^{21}+b_{12}^{22}+b_{12}^{11}+b_{12}^{12}\right) & \cdots \\
        \frac{1}{2}\left(b_{21}^{21}+b_{21}^{22}+b_{21}^{11}+b_{21} ^{12}\right) & -1+\frac{1}{2}(b_{22}^{21}-b_{2}^{12})-\lambda & \cdots \\ \vdots & \vdots & \ddots 
    \end{bmatrix}
\end{align}
Using Wigner's circle law, eigenvalues that solve $\det(\lim_{\Delta\to\infty}(\mathbf{W-XZ^{-1}Y}))=0$ will be distributed in a circle centered at $-1$ with radius $\sigma\sqrt{SC}$, thus returning us to the classical stability limit of $K<1$.
}

\ed{Repeating the procedure for the second set of eigenvalues, 
\begin{align}
    \mathbf{W}=\begin{bmatrix}
        -1+b_{11}^{12}-\lambda & b_{12}^{11}+b_{12}^{12}& \cdots  \\b_{21}^{11}+b_{21}^{12} & -1+b_{21}^{12} -\lambda& \cdots \\ \vdots & \vdots & \ddots 
    \end{bmatrix} 
    = (-1-\lambda)\Id + \underbrace{\begin{bmatrix}
        b_{11}^{12} & b_{12}^{11}+b_{12}^{12}& \cdots  \\b_{21}^{11}+b_{21}^{12} & b_{21}^{12} & \cdots \\ \vdots & \vdots & \ddots 
    \end{bmatrix}}_{\mathbf{N}}
\end{align}
At large $\Delta$,
\begin{align}
    \mathbf{W}^{-1} = 
        \frac{-1}{1+\lambda}\Id
\end{align}
Then \begin{align}
    \mathbf{Z-YW^{-1}X} = &\begin{bmatrix}
        -1-b_{11}^{12}-2\Delta -\lambda & b_{12}^{22}-b_{12}^{12} & \cdots \\
        b_{21}^{22}-b_{21}^{12} & -1-b_{22}^{12}-2\Delta-\lambda & \cdots \\
        \vdots & \vdots & \ddots 
    \end{bmatrix} \nonumber\\ 
    &+ \frac{1}{1+\lambda}\begin{bmatrix}
        b_{11}^{21}-b_{11}^{12} & b_{12}^{21}+b_{12}^{22}-b_{12}^{11}-b_{12}^{12} & \cdots \\ b_{21}^{21}+b_{21}^{22}-b_{21}^{11}-b_{21}^{12} & b_{22}^{21}-b_{22}^{12} & \cdots \\ \vdots & \vdots & \ddots 
    \end{bmatrix}\begin{bmatrix}
        b_{11}^{12}+\Delta & b_{12}^{12}& \cdots \\ b_{21}^{12} & b_{22}^{12}+\Delta & \cdots \\ \vdots & \vdots & \ddots 
    \end{bmatrix}
\end{align}
\begin{align}
    \lim_{\Delta\to\infty}(\mathbf{Z-YW^{-1}X}) = &\begin{bmatrix}
        -1-b_{11}^{12}-2\Delta -\lambda & b_{12}^{22}-b_{12}^{12} & \cdots \\
        b_{21}^{22}-b_{21}^{12} & -1-b_{22}^{12}-2\Delta-\lambda & \cdots \\
        \vdots & \vdots & \ddots 
    \end{bmatrix} \nonumber\\
    &+\frac{\Delta}{1+\lambda}\begin{bmatrix}
        b_{11}^{21}-b_{11}^{12} & b_{12}^{21}+b_{12}^{22}-b_{12}^{11}-b_{12}^{12} & \cdots \\ b_{21}^{21}+b_{21}^{22}-b_{21}^{11}-b_{21}^{12} & b_{22}^{21}-b_{22}^{12} & \cdots \\ \vdots & \vdots & \ddots 
    \end{bmatrix}
\end{align}
Where in the last line we have dropped terms smaller than order $\mathcal{O}(\Delta)$ from the second term.
}

\ed{Again eigenvalues will be near $-2\Delta$, so we take $\frac{\Delta}{1+\lambda}\to-\frac{1}{2}$, and \begin{align}
    \lim_{\Delta\to\infty}(\mathbf{Z-YW^{-1}X}) = \begin{bmatrix}
        -1-2\Delta -\frac{1}{2}(b_{11}^{12}+b_{11}^{21})-\lambda & \frac{1}{2}(b_{12}^{11}+b_{12}^{22}-b_{12}^{12}-b_{12}^{21}) & \cdots \\
        \frac{1}{2}(b_{21}^{11}+b_{21}^{22}-b_{21}^{12}-b_{21}^{21}) & -1-2\Delta -\frac{1}{2}(b_{22}^{12}+b_{22}^{21})-\lambda & \cdots \\ \vdots & \vdots & \ddots 
    \end{bmatrix}
\end{align}
This again has eigenvalues near $-2\Delta$ distributed with  radius $\sigma\sqrt{SC}.$
}

\ed{Thus, stability criteria only depends on the group near the origin and we return to the classical stability criteria of $K<1$ required for a stable system.}

\subsection*{Interpretation of strongly-negative intra-species interactions}

In the case where stages have independent interactions ($\rho=0$), we can reformulate to include the quadratic intra-species interactions separately:\begin{equation}
    \frac{dx_i^\alpha}{dt}=\sum_{j,\beta} R_{ij}^{\alpha\beta}x_j^{\beta}+x_i^\alpha\left( \sum_{j,\beta} B_{ij}^{\alpha\beta}x_j^\beta+ \sum_j Q_{ij}^{\alpha\beta}x_j^\beta\right)
\end{equation} where $\Bind$ with random intra-species interactions is used, and \begin{equation}
    \Qq = \begin{bmatrix}
        0 & q^{\phi}_{1} \\ q^{\gamma}_{1} & 0 \\ & & 0 & q^{\phi}_2 \\& & q^{\gamma}_2 & 0 \\ & & & & \ddots
    \end{bmatrix}
\end{equation}
If elements in $\Qq $ are $-1$ or negative, $\Bind + \Qq \sim \Bb_{\rho=0}$. This can be interpreted as having a quadratic term in $\Rr$, where in the two-stage case: \begin{align}
    \frac{dx_i^1}{dt} &= \mu_i^1 x_i^1 + \left(\phi_i + q^{\phi}_ix_i^1\right)x_i^2 +x_i^1\sum_{j,\beta}B_{ij}^{1\beta}x_j^\beta\\
    \frac{dx_i^2}{dt} &= \left(\gamma_i + q^\gamma_{i}x_i^2\right)x_i^1 + \mu_i^2x_i^2 +x_i^2\sum_{j,\beta}B_{ij}^{2\beta}x_j^v
\end{align}
The quadratic terms added to $\Rr$ effectively create new effective growth and fecundity rates: \begin{align}
    \phi_{i}' = \phi_i + q_i^\phi x_i^1 \qquad \gamma_i'=\gamma_i+q_1^\gamma x_i^2
\end{align} The growth rate of juveniles changes dynamically with the current adult population, and likewise the adult fecundity changes with changes in the density of juveniles. When $q^\phi,\ q^\gamma < 0$, this is a stabilizing effect, and simply a reinterpretation of the competitive intra-specific interactions in our $\Bb_\rho$ cases, similar to the mechanisms suggested by \citet{deRoos2021}.

\ed{
\subsection*{Effect of connectivity in the interaction matrix}
When drawing entries for the interaction matrix, entries are nonzero with probability $C$ and then drawn from the Normal distribution $\mathcal{N}(0, \sigma^2)$. We take $C=1$ in every case to focus on the effect of changing the variance of interaction strengths $\sigma^2$.}
\ed{
\subsubsection*{Single-stage case with $\Aa$}
In the single-stage case with $S$ species, interaction matrix $A$ has $S^2-S$ off-diagonal entries drawn from $\mathcal{N}(0,\sigma^2)$. When $C\neq 1$, the probability that each entry is drawn from this distribution is $C$. Thus $a_{ij}$ is drawn from one of two distributions: 
\begin{align}
    a_{ij} = 
    \begin{cases}
        x\sim N(0,\sigma^2) & \text{with probability } C \\
        y=0 &\text{with probability } 1-C
    \end{cases}
\end{align} The expectation of $a_{ij}$ is still $E[a]=0$. The variance of any random variable is defined as \begin{align}
    \text{Var}(Z)=E[Z^2]-(E[Z])^2
\end{align} The expected value of the squared variable is \begin{align}
    E[a^2]=C\cdot \text{Var}(x)^2 + (1-C)\cdot Var(y)^2 = C\sigma^2
\end{align} and \begin{align}
    Var(a) = C\sigma^2
\end{align}
The complexity of the community is defined as $K=\sqrt{\sigma^2 S C}= \sqrt{S\cdot Var(a)}$, and so only depends on $C $ in how the connectivity affects the variance of off diagonal elements in $A$. Thus, there is no explicit dependence on $C$ and we can set it to 1 to focus specifically on changing $K$ via $\sigma$.
}

\ed{
\subsubsection*{Multiple life stages, with $\Bb$}
With multiple life stages using $\Bb_{\rho}$, $\Bb$ has $S^2-S$, $L\times L$ blocks. Each block is nonzero with probability $C$, and if so has entries drawn from $b\sim N(0, \sigma^2)$ with covariance $\rho\sigma^2$. Thus we again will see that the overall variance of entries in $\Bb$ is $\text{Var}(b)=C\sigma^2$. This is verified numerically in Figure \ref{fig:ap_C_2stage}.
}
\ed{
\begin{figure}[!htb]
\includegraphics[width=1\textwidth]{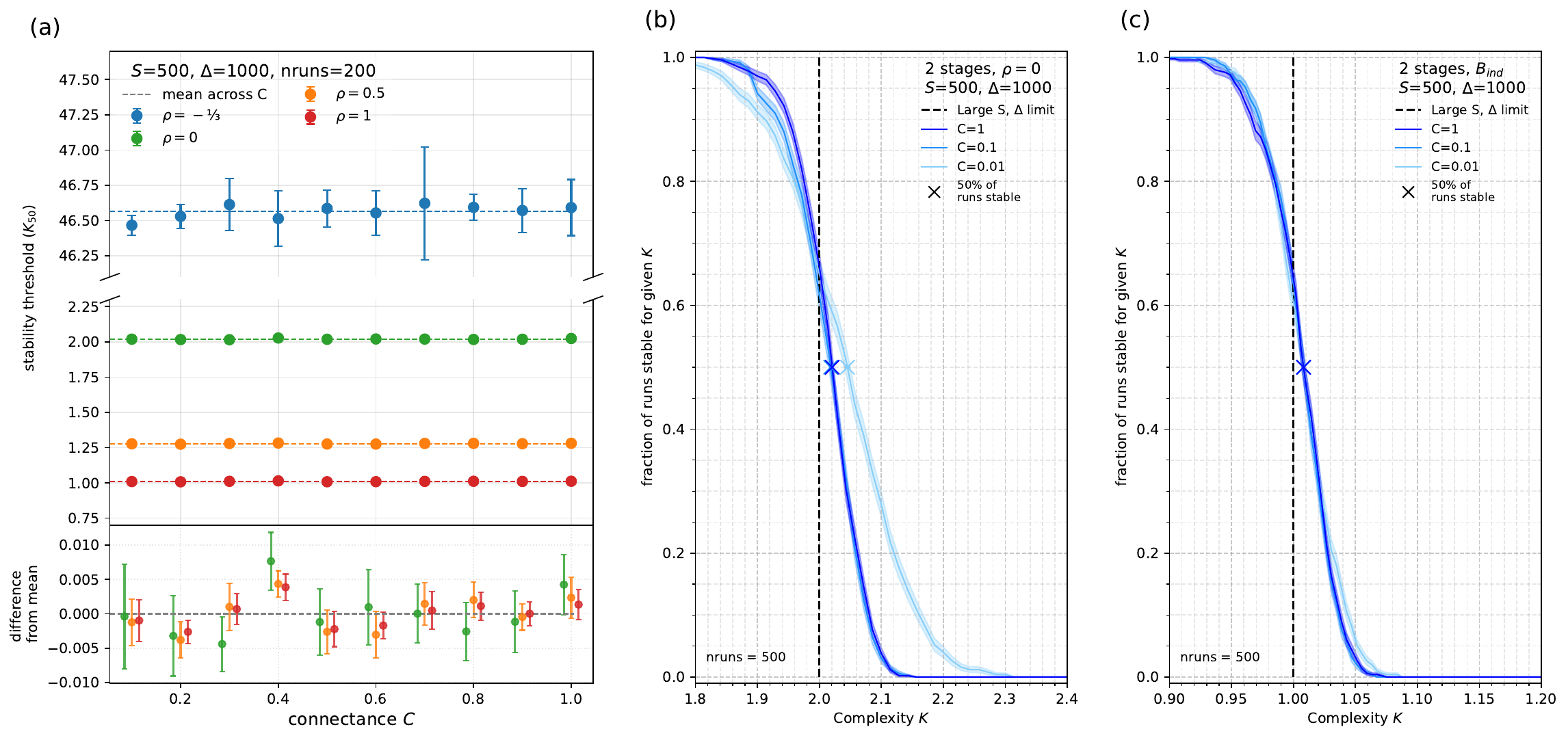}
\caption{\label{fig:ap_C_2stage} 
\ed{For our two stage cases, we ensure robustness to differen values of connectedness $C$ numerically. In (a), we show $K_{50}$, the value of $K$ that gives $50\%$ of random realizations as stable, for different $\rho$ in $\Bb_\rho$. The dashed line shows the mean across the 10 different values of $C$ tested. For $\rho=0, 0.5, 1$ errors away from the mean are shown in the bottom panel, as they are smaller than the markers in the upper panel. Although the variation of $K_{50}$ across $C$ increases with decreasing $\rho$, for the full spectrum of $\rho$ the value of $C$ has no significant impact on the value of $K_{50}$ measured across different realizations. In (b) and (c), we measure the fraction of stable runs over 500 random realizations per point, across a range of $K$ for $\Bb_{\rho=0}$(b) and $\Bb_{ind}$(c). For decreased $C$, we see a similar effect as in smaller system size (reduced $S$), but the qualitative behavior across the range of $K$ tested remains the same. All errors shown come from bootstrapping across the indicated number of nruns.
}}
\end{figure}
}

\ed{In the case of $B_{ind}$, the interaction matrix is identical to a single stage matrix with $2S$ species, and thus the same procedure applies as in the single stage case.
}

\ed{\subsection*{3 or more life stages}
Our interaction matrix with life stages $\Bb$ can be extended beyond 2 stages, but the form of $\Rr$ must be considered. If we consider for species $i$ that each stage $\alpha$ for $\alpha<L$ can only mature into the next stage with rate $\gamma_i^\alpha$, and only the final stage $\alpha=L$ can reproduce the first stage with rate $\phi_i$, We will have $2L$ nonzero parameters in $\Rr$ for each species. In the 3-stage case, $\Rr$ will be:
\begin{align}
    \Rr_{(L=3)} = \begin{bmatrix}
        \mu_1^1 & 0 & \phi_1 \\
        \gamma_1^1 & \mu_1^2  & 0 \\
        0 & \gamma_1^2 &\mu_1^3 \\
        & & & \mu_2^1 & 0 & \phi_2 \\
        & & & \gamma_2^1 & \mu_2^2  & 0 \\
        & & & 0 & \gamma_2^2 &\mu_2^3 \\
        & & & & & & \ddots 
    \end{bmatrix}
\end{align}
Where all entries outside the $L\times L$ diagonal blocks are 0.
}

\ed{Similarly, for 4 stages we will have
\begin{align}
    \Rr_{(L=4)} = \begin{bmatrix}
        \mu_1^1 & 0 & 0 & \phi_1 \\
        \gamma_1^1 & \mu_1^2  & 0 & 0 \\
        0 & \gamma_1^2 &\mu_1^3 & 0 \\
        0 & 0 & \gamma_1^3 &\mu_1^4 \\
        & & & &\mu_2^1 & 0 & 0 & \phi_2 \\
        & & & &\gamma_2^1 & \mu_2^2  & 0 & 0 \\
        & & & &0 & \gamma_2^2 &\mu_2^3 & 0 \\
        & & & &0 & 0 & \gamma_2^3 &\mu_2^4 \\
        & & & & & & & & \ddots 
    \end{bmatrix}
\end{align}
Following a similar procedure as in the 2-stage case with some value $\Delta$ to set demographic rates (see \eqref{M-eq:Rvalmins}in main text), we get: \begin{align}
    \phi_i = \text{max}\left(0, -\sum_{j,\beta}b_{ij}^{\alpha\beta}\right)+\Delta
\end{align}
\begin{align}
    \gamma_i^\alpha=\text{max}\left(0, -\sum_{j,\beta}b_{ij}^{(\alpha+1)\beta}\right)+\Delta \quad \text{for }i=1:L-1
\end{align}
\begin{align}
    \mu_i^1 = -\sum_{j,\beta}b_{ij}^{1\beta} -\phi_i 
\end{align}
\begin{align}
    \mu_i^{\alpha} = -\sum_{j,\beta}b_{ij}^{\alpha\beta}-\gamma_i^\beta
 \quad \text{for } i=2:L
 \end{align}
Using these $\Rr$ matrices with interaction matrices generalized from \eqref{M-eq:Brhomatrix} and \eqref{M-eq:Bindmatrix}, we numerically verify that decreased $\rho$ continues to increase the boundary of asymptotic stability in the 3-stage case, as shown in fig \ref{fig:3stage}.
}

\ed{\begin{figure}[!htb]
\includegraphics[width=1\textwidth]{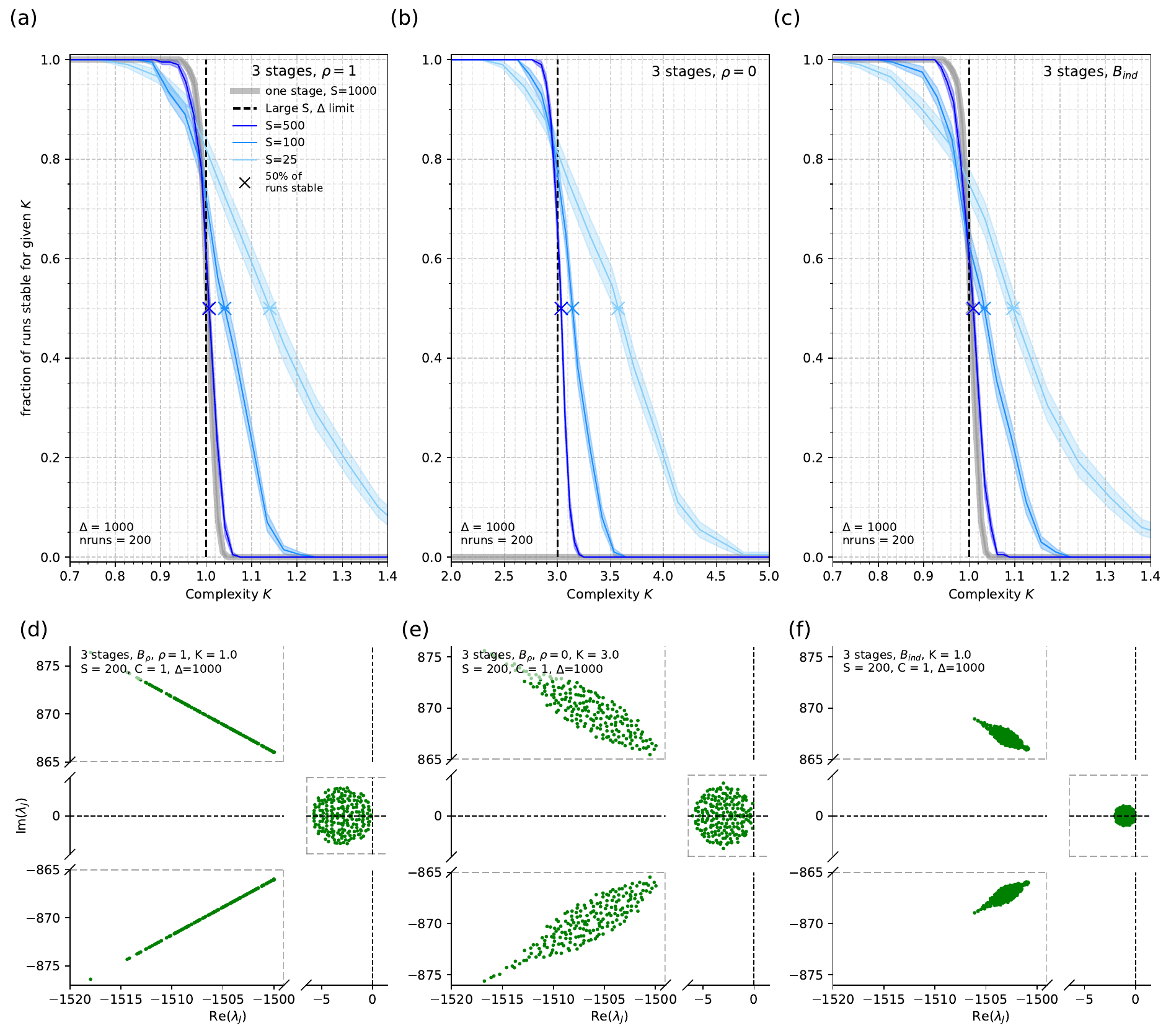}
\caption{\label{fig:3stage} 
\ed{For each of the $\rho=1$, $\rho=0$, and $\Bb_{ind}$ cases with 3 stages, the fraction of stable configurations given a specified complexity $K$ is computed (a, b, c). At $S=500$, the value of $K$ for which 50\% of runs are stable approaches the asymptotic value. From our numerical analysis, the transition to instability occurs at $K=1$ for the $\rho=1$ case and the case with random intra-specific intractions (a, b), but is increased beyond the classical threshold for $\rho=0$ (c). The eigenvalue distributions for the three cases are also shown (d, e, f) and again we see $S$ eigenvalues distributed in a circle determining stability for each case, and $2S$ eigenvalues are far from the $Re(\lambda_J)=0$ boundary for asymptotic instability. }
}
\end{figure}}

\ed{\subsection*{Numerical Methods}
Figures and results for finite values of $S$ and $\Delta$ are generated numerically, using the following described algorithms. Code is available at \url{https://github.com/ggarmire/lvlh_final_final/}.
}
\ed{
\subsubsection*{Matrix generation}
For each simulation we perform, we must generate an interaction matrix ($\Aa$ or $\Bb$) and a vector of growth rates ($\vec{r}$) or demographic matrix ($\Rr$). For each case, we control the random number generation by setting the seed for reproducibility.}

\ed{Each version of the interaction matrix is produced as follows: 
\begin{itemize}
    \item \textbf{Single-stage case $\Aa$:} To model single-stage dynamics, we construct an $S\times S$ matrix following previous works \citep{may1972, Allesina_2012}. Intra-specific competition is enforced by setting diagonal entries $a_{ii}$ to $1$. Off-diagonal entries, for inter-species interactions, are assigned to be nonzero with connectance $C$. Nonzero entries are independent and identically distributed and drawn from the normal distribution with mean $0$ and variance $\sigma^2$. Therefore entries of $\Aa$ follow: 
    \begin{align}
        a_{ij} = \begin{cases}
            -1 & \text{if }i=j \\
            N(0, \sigma^2) & \text{if }i\neq j \text{ with probability } C\\
            0& \text{if }i\neq j\text{ with probability } 1-C
        \end{cases}
    \end{align}
    matching the form used in previous works \eqref{M-eq:Amatrix} \citep{may1972, Allesina_2012} 
    \item \textbf{Stage-dependent interactions $\Bb_\rho$}: For the stage-structured community of $S$ species each with $L$ stages, we construct and $LS\times LS $ matrix composed of $L\times L$ sub-matrices $B_{ij}$:\begin{align}
        \Bb = \begin{bmatrix}
            \Bb_{11} & \Bb_{12} & \Bb_{13} & \cdots \\
            \Bb_{21} & \Bb_{22} & \Bb_{23} & \cdots \\
            \Bb_{31} & \Bb_{32} & \Bb_{33} & \cdots \\
            \vdots & \vdots & \vdots & \ddots 
        \end{bmatrix}
    \end{align}Intra-specific competition is again enforced independently of stage by filling diagonal blocks $\Bb_{ii}$ to $-1$.  Off-diagonal sub-matrices are set to $0$ with probability $1-C$. With probability $C$, the $L^2$ elements of a sub-matrix $\Bb_{ij}$ are drawn from a multivariate normal distribution with mean $0$ and coviariance matrix $\Sigma$. To account for the correlation of inter-species interactions of different life stages, the covariance matrix is defined as \begin{align}
        \Sigma = \sigma^2\left(\rho \mathbf{1}+(1-\rho)\Id\right)
    \end{align}
    where $\mathbf{1}$ is the $L\times L$ matrix of $1$ and $\Id$ is the $L\times L$ identity. Therefore blocks of $\Bb_\rho$ follow \begin{align}
        \Bb_{ij}=\begin{cases}
            -1_{L\times L} & \text{if }i=j \\
            N(0, \Sigma) & \text{if } i\neq j \text{ with probability } C\\
            0_{L\times L} & \text{if } i\neq j \text{ with probability } 1-C
        \end{cases}
    \end{align} 
    For $\rho=1$, stages are maximally correlated and interact identically, so each block $\Bb_{ij}$ is filled with $L^2$ identical entries. For $\rho=0$, nonzero blocks $\Bb_{ij}$ have $L^2$ independent entries. For a system with  $L$ stages per species, $\rho$ is minimally bounded by $\rho\geq \frac{1}{L^2-1}$, which is the maximum anti-correlation between $L^2$ variables. 
    \item \textbf{Random intra-specific interactions $\Bb_{ind}$:} When intra-specific interactions are not fixed to be competitive between life stages, we draw all off-diagonal entries of $\Bb_{ind}$ independently from the normal distribution with mean $0$ and variance $\sigma^2$. Thus $\Bb_{ind}$ is drawn with the same method as the single-stage case, but for a  $2S\times 2S$ matrix. Each interaction term representing the effect on species $i$ stage $u$ of species $j$ stage $v$ is: 
    \begin{align}
        b_{ij}^{\alpha\beta} = \begin{cases}
            N(0, \sigma^2)_{_{_{_{_{_{_{_{_{}}}}}}}}} & \text{if }i= j\text{ and } {\alpha\neq \beta} \text{, with probability } C\\ 
            N(0, \sigma^2) & \text{if }i\neq j \text{, with probability } C\\
            0& \text{if }i\neq j\text{, with probability } 1-C
        \end{cases}
    \end{align}
    \end{itemize}
}
\ed{In each case, we scale the standard deviation of our interaction strengths as $\sigma = K(SC)^{-0.5}$ so that we may vary $K$ for each model independently of system size. Because this scaling is applied identically in generating both $S\times S$ and  $LS\times LS$ interaction matrices, the overall variance remains mathematically constant across models. Thus we can evaluate both model types at the same $K$ values for direct comparison. }

\ed{Demographic rates are determined using one of the two following methods:
\begin{itemize}
    \item In the single-stage case, the growth vector $\vec{r}$ if fully constrained by the entries of $\Aa$ and the feasibility condition requiring $\vec{x}^*=\vec{1}$. For each species $i$, the growth rate is 
\begin{align}
    r_i=-\sum_j a_{ij}
\end{align} which satisfies $0=r_ix^*_i+\sum_{j}A_{ij}x^*_j$ for the equilibrium at $\vec{x}^*=1$.
    \item In the two stage case, rates in $\Rr$ are under-constrained. For a feasible equilibrium at $\vec{x}^*=\vec{1}$ in the two-stage case, we require
\begin{align}
    0 =  \mu_i^1 + \phi_i + \sum_{j,v}b_{ij}^{1v}\\
    0 =  \gamma_i + \mu^2_i + \sum_{j,v}b_{ij}^{2v}
\end{align}
Elements of $\Rr$ are thus drawn as 
\begin{align}
    \phi_i=\text{max}\left(0, -\sum_{j,v}b_{ij}^{1v}\right)+\Delta &\qquad  \gamma_i=\text{max}\left(0, -\sum_{j,v}b_{ij}^{2v}\right)+\Delta \\\nonumber \\
    \mu_i^1 = -\sum_{j,v}b_{ij}^{1v} -\phi_i &\qquad \mu_i^2 = -\sum_{j,v}b_{ij}^{2v} -\gamma_i
\end{align}
Thus all elements in $\Rr$ are constrained by the interaction matrix and a choice of parameter $\Delta$, which sets the minimum magnitude of any of these demographic rates.
\end{itemize}}

\ed{
\subsubsection*{Single simulation}
To simulate the reaction of one community to a perturbation near its feasible equilibrium, we use our parameters to generate some interaction matrix and demographic rates, and evolve the system forward in time from an initial condition near $\vec{x}^*=\vec{1}$. We implement this computationally through the following framework: 
\begin{itemize}
    \item \textbf{Matrix generation:} We construct the interaction matrix $\Aa$ or $\Bb$ following the appropriate procedure above. Then $\vec{r}$ is calculated using \eqref{M-eq:onestageconstraint} for the single stage case, or \eqref{M-eq:2stagecontraint} is used with some chosen $\Delta$ to construct $\Rr$.
    \item \textbf{Initial condition:} The initial condition is used to simulate a perturbation away from the known equilibrium of $\vec{x}^*=\vec{1}$. Thus $\vec{x}(0)$ is a vector of length $L\cdot S$ with entries selected from a uniform distribution centered at $1$.
    \item\textbf{Numerical integration:} The system of non-linear ordinary differential equations (from \eqref{M-gLV} or \eqref{M-eq:2stagedtnamics}) are solved numerically using the \texttt{scipy.integrate.odeint} module.
    \item\textbf{Extinction handling:} To prevent numerical noise from creating unrealistic behaviors, such as asymptotically small negative population densities, we enforce an extinction threshold inside the solver function. If any sub-population falls below $10^{-8}$, we reset the density to 0 and consider it extinct. 
    \item\textbf{Stability evaluation:} By observing the population trajectories over time, we can see them converge or diverge respectively for a stable or unstable system. Stability is confirmed via analysis of the leading eigenvalue.
\end{itemize}
Some examples of evolved communities are shown in Figure \ref{fig:ap_systemevolutio}.
\begin{figure}[!htb]
\includegraphics[width=1\textwidth]{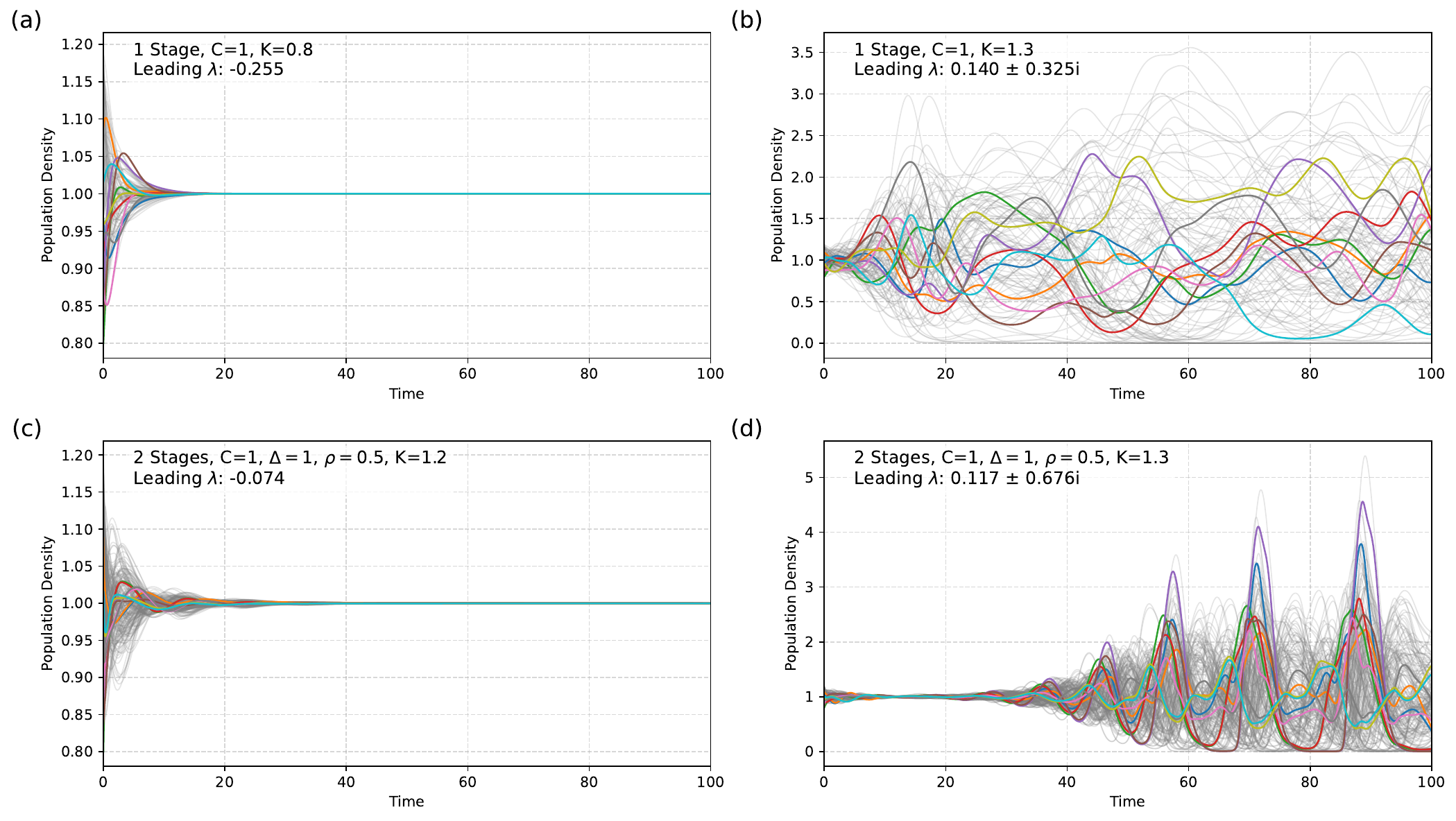}
\caption{\label{fig:ap_systemevolutio} 
\ed{The evolution of 4 different communities is found by solving the nonlinear ordinary differential equations numerically. In (a) and (b), a single stage case with 100 species is shown in (a) an unstable ($K=0.8$)  and (b) a stable ($K=1.3$) configuration set by the interaction matrix. In the stable configuration, populations return to their equilibrium densities of $1$, while in the unstable configuration a small perturbation causes populations to diverge from $1$ and some species go extinc. In (c) and (d), a system of 100 species, each with 2 stages, is shown with a value of $\rho=0.5$ used to select interactions.  Again (c) a stable configuration ($K=1.2$) sees the species return to equilibrium after a perturbation, while (d) in the unstable configuration ($K=1.3$) species populations oscillate with increasing amplitude and begin to go extinct within the simulated timeframe. }
}
\end{figure}
}

\ed{
\subsubsection*{Transition to Instability}
We visualize the transition to instability for a given set of parameters, as in Figure \ref{M-fig:Scurves}, numerically by calculating the fraction of realizations of an interaction matrix that result in a Jacobian at equilibrium with all negative eigenvalues. For some given $S$, $\Delta$ and form of interaction matrix $\Bb$, we construct an ensemble of independent matrices utilizing our seed setting. To map out the complete stability phase space, we sweep through a defined range of complexity values, $K$, conducting a fixed number of independent network realizations (typically 200 runs) at each incremental step.
}

\ed{
For each realization, we construct the corresponding $\Rr$ matrix and evaluate the Jacobian at equilibrium. A specific realization is classified as stable if the real part of the leading eigenvalue is negative. Because evaluating hundreds of high-dimensional matrices across many values of $K$ is computationally intensive, these tasks are heavily parallelized. At each value of $K$, the fraction of stable runs is calculated as the mean of the binary stability outcomes. A bootstrapping algorithm is used at each point to calculate the standard error displayed on figures by the shaded regions.
}

\ed{
To find $K_{50}$, or the value of $K$ resulting in $50\%$ of runs stable, we implement an automated search algorithm. The procedure begins at an initial guess for $K$ and calculates the fraction of stable runs, $f(K)$, for some number of runs (typically set to 200 runs). Based on the result, the algorithm steps forward or backward in $K$ until it successfully brackets the $50\%$ threshold by finding adjacent $K$ values for which $f(K_{low})<0.5$ and $f(K_{high})>0.5$. Then the bounds are narrowed by iteratively evaluating the midpoint, until both bracketing bounds are between $0.4$ and $0.6$. Finally a linear interpolation is performed to extract the estimate of $K_{50}$. The statistical uncertainty is found by bootstrapping the binary (stable/unstable) outcomes of the simulations at the final bracketing $K$ values.
}

\begin{figure}[!htb]
\includegraphics[width=1\textwidth]{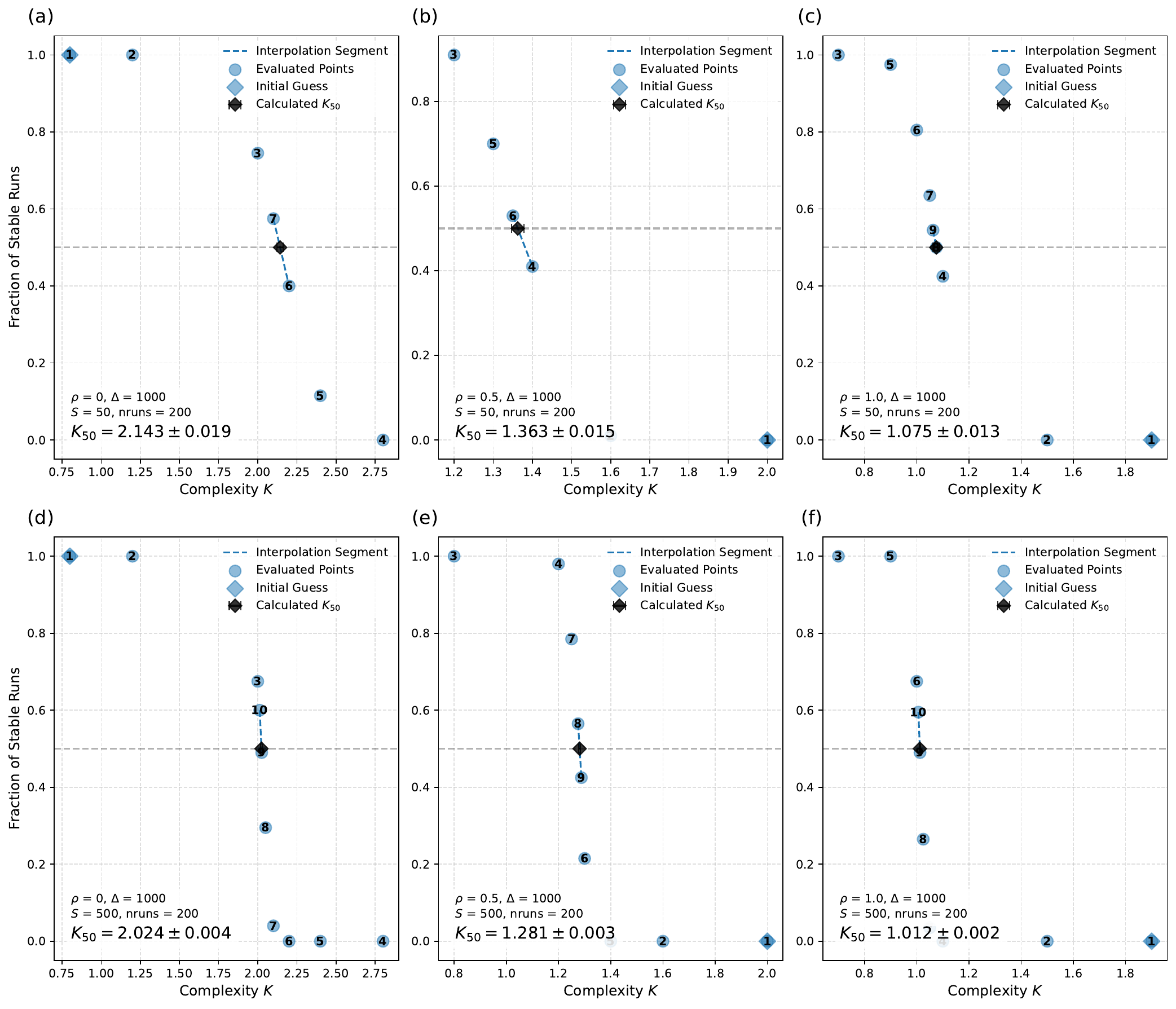}
\caption{\label{fig:ap_K50search} 
\ed{Numerical algorithm for finding $K_{50}$, the value of $K$ for which $50\%$ of random realizations of the interaction matrix are predicted to be stable. We start with some initial guess of $K$, marked as point 1. The algorithm iteratively steps through values of $K$ until two points on either side of the $50\%$ stability threshold can be interpolated between, shown by the dashed line. The estimated $K_{50}$ is shown by the black diamond with its error found by bootstrapping across the nruns used in interpolation. Points are labeled in the order they are found via our algorithm, and steps are shown for two different values of $S$ (50, 500) across three different values of $\rho$ (0, 0.5, 1). An identical procedure is used for cases with $\Bb_{ind}$.}
}
\end{figure}

\ed{
Points in Figure \ref{M-fig:deltafig} are found with the algorithm described above for different values of $S$ and $\Delta$, and different forms of interaction matrix $\Bb$.
}

\ed{
\subsection*{Estimation of uncertainties }
To quantify uncertainty in our numerical simulations,  the standard error is calculated with a bootstrapping algorithm. This is done slightly differently for the stable fraction $f(K)$ and when interpolating to find $K_{50}$. 
}

\ed{
When estimating the stable fraction using at some specific  value of $K$, as each point in Figure \ref{M-fig:Scurves}, we simulate many independent realizations of the interaction matrix (typically $200$ runs). and record the binary stability outcomes of each. To estimate the standard error of the mean stable fraction, we perform $1000$ resampling iterations. In each iteration, we draw a new sample of identical size from the original outcomes with replacement and calculate its mean. The standard error is then defined as the sample standard deviation of these $1000$ bootstrapped means.
}

\ed{
Estimating the error on the critical threshold $K_{50}$ requires a modified bootstrapping approach applied directly in the interpolation step. We isolate the binary stability outcomes of the two $K$ values that bracket $K_{50}$, denoted as $K_{low}$ and $K_{high}$. We then generate $1000$ resampling indices and apply these to both the $K_{low}$ and $K_{high}$ outcomes. This gives $1000$ pairs of bootstrapped fractions, $f_{low}$ and $f_{high}$. For each independent iteration, we calculate a bootstrapped threshold by re-interpolating using each pair of $(f_{low}, f_{high})$. The final standard error is found as the standard deviation of the resulting distribution of $1000$ interpolated values. 
}

\pagebreak
\bibliographystyle{ecology_letters2}
\bibliography{bib/gracerefs}{}